%% file: kummer_schutz_v03.tex
\documentclass[a4paper]{article}

\usepackage{latexsym}
\usepackage{amsthm}
\usepackage{amssymb}
\usepackage{amsfonts}
\usepackage{amsmath}
\include{smac}

\newcommand{\oQ}{{}``}
\newcommand{\eQ}{''}
\begin{document}

\setlength{\parindent}{0mm}
\bigskip{}
{\centering 
\textbf{\large LINEAR DERIVATIVE CARTAN FORMULATION OF GENERAL RELATIVITY}\large
\par}

\bigskip{}
{\centering W.\ Kummer and H.\ Sch\"utz\par}

\medskip{}
{\centering \textit{Institute f.\ Theor.\ Physics, Vienna University of Technology}
\par}

{\centering \textit{Wiedner Hauptstrasse 8-10, A-1040 Vienna, Austria}\par}

\begin{abstract}
\bigskip{}
\noindent 
Beside diffeomorphism invariance also manifest SO(3,1) local Lorentz invariance
is implemented in a formulation of Einstein Gravity (with or without
cosmological term) in terms of initially  completely independent vielbein
and spin connection variables and auxiliary two-form fields.
In the systematic study of all
possible embeddings of Einstein gravity into that formulation with 
auxiliary fields, the introduction of a ``bi-complex'' algebra possesses
crucial technical advantages. 
Certain components of the new two-form fields
directly provide canonical momenta for spatial components of all Cartan
variables, whereas
the remaining ones act as Lagrange multipliers for a
large number of constraints, some of which have been proposed already
in different, less radical approaches. The time-like components of the
Cartan variables play that role for the Lorentz constraints and others 
associated to the vierbein fields. Although also some ternary 
ones appear, 
we show that relations exist between these constraints, and how the
Lagrange multipliers are to be determined to take care of second 
class ones. We believe that our  
formulation of standard Einstein gravity as a gauge theory with consistent 
local Poincar\'e algebra is superior to earlier similar attempts.

\end{abstract}

\section{Introduction}

For many decades the standard approach to the Hamiltonian formulation
of General Relativity (GR) has been the use of specific geometric
variables, introduced by Arnowitt, Deser and Misner (ADM) \cite{ADM62}:
the lapse and shift functions, the intrinsic curvature
and the three space metric on  space-like surfaces which determine
the foliation corresponding to a time coordinate \( x^{0} \).
 The Hamiltonian analysis
leads to four constraints, the Hamiltonian and the diffeomorphism 
ones, which are all first class. 
At the quantum level the Hamiltonian constraint
can by transcribed as a formal functional differential equation, the so-called
Wheeler-DeWitt equation \cite{DEW67}.

From the point of view of geometry it is well known that the description
of a manifold by Cartan variables \cite{CART28}, (one-form) vierbeine
\( e^{A} \) and (one-form) spin connection \( \omega ^{AB} \) (with
\( A=0,1,2,3 \), describing the local Lorentz coordinates), is the
most comprehensive one. Only when the torsion vanishes, as it is assumed
in Einstein's General Relativity (GR), the spin connection is expressed in terms of the vierbeine
and ceases to represent an independent variable. In this interpretation
the metric \( g=\eta ^{AB}e_{A}\otimes e_{B} \) \( (\mathrm{diag}\, 
\eta =(-1,1,1,1)) \)
is  not a fundamental, but rather a derived quantity. It owes its 
existence to the appearence of just this combination of vierbeine 
in a Lagrangian of a scalar test particle.

The opposite interpretation consists in taking the spin connection
as the more basic variable.\footnote{The most consequent
{}``connection'' formulation of gravity is perhaps the one by
Capovilla and Jacobson \cite{CDJ91}.}
In that way a closer resemblance to the
well-studied Yang-Mills theories could be achieved. This has been
the key to Ashtekar's identification of an (anti-)self dual
spin connection with an SO(3) Yang-Mills field \cite{ASHT86}
which by introducing notions  from the Yang-Mills case
like Wilson loops allowed many important new developments (cf 
e.g.\  \cite{ROSMO}).  The price to pay in
Ashtekar's original formulation was the extension to complex
fields which,
in the end, had to be reduced by imposing a reality condition.
Including a second SO(3) connection proportional to the Immirzi parameter
\( \beta  \),  Barbero \cite{BARBERO} showed that most of
the advantages of Ashtekar's formulation (e.g. polynomial constraints)
could be retained for a real value of \( \beta  \) which in Ashtekar's
approach would have to be imaginary.
However, that parameter  
introduced an ambiguity (``Immirzi ambiguity'') in a quantized theory
 \cite{ROTH98}. At present perhaps the most
elaborate version of this general line is the one pursued up to the
level of quantum theory by T. Thiemann \cite{THIE96}. 

Still, also in the developments originating from the connection type
formulation the vanishing torsion condition is incorporated as a given 
relation between connection and vierbeine from the start. 

On the other hand, the idea to at least initially treat 
both types of variables
as independent ones, goes back to Palatini \cite{PALA19}%
\footnote{More precisely, he viewed the metric and the covariant 
derivative as independent variables in the Hilbert-Einstein action.} 
who showed it to be a peculiar property of the Hilbert-Einstein action
for GR that it also yields the metricity condition and the condition
of vanishing torsion. In its modern version \cite{KIBB61} in
terms of Cartan variables the {}``Hilbert-Palatini'' action
$(\epsilon_{0123}=-\epsilon^{0123}=1)$\footnote{This follows directly by
 interpreting $\epsilon$ as volume form and following the usual index 
convention $f_A := f(E_A)$ for indices $A,B \ldots$ from the 
begining of the alphabet, referring to local (anholonomic) 
Lorentz coordinates $A=(\ul{0},\ul{1},\ul{2},\ul{3})=(\ul{0},a)$. 
Indices $I=(0,1,2,3)=(0,i)$ from 
the middle of the alphabet are related to holonomic coordinates. 
In the action the prefactor, depending on Newton's constant, is 
normalized to one.}, 
\begin{equation}
\label{ActHP}
S_{(HP)}=\frac{1}{2}\int\limits _{\mathcal{M}_{4}}R^{AB}
\wedge e^{C}\wedge e^{D}\epsilon _{ABCD}\;,
\end{equation}
with the curvature two-form 
\begin{equation}
\label{DefR}
R^{AB}=
d\omega^{AB} + {\omega^{A}}_{C} \wedge \omega^{CB}\; ,
\end{equation}
by independent variation of the vierbeine \( e_{A} \) and of the spin
connection $\omega_{AB}$ fields yields vanishing torsion
\begin{equation}
\label{DefTor}
\tau ^{A}=\left( De\right) ^{A}=de^{A}+\omega ^{A}_{\, \, B}\wedge e^{B} = 0
\end{equation}
 and the Einstein equations of GR. 
Because (\ref{ActHP}) only contains 
first derivatives of \( \omega ^{AB} \), this is usually referred to as the 
first order formulation of GR. 

The connection form can be regarded as an SO(3,1) gauge field. This
implies that \( \omega ^{AB} \) defines a connection in a principal
fiber bundle \( \mathcal{P}\, \left( \mathcal{M}, SO( 3,1) \right)  \)
or vector bundles associated with it. However, there is a basic difference
to ordinary gauge theories: the local gauge symmetry is not independent
from the effective action on the manifold, it is rather the same.
Nevertheless, the interpretation as a gauge field turns out to be
more successful when \( \omega  \) is treated as an independent field.

When the description of geometry by Cartan variables, instead of the
metric as in the original ADM formalism, is taken seriously it is
natural to include Lorentz constraints \( \Omega ^{AB} \) which satisfy
the Poisson brackets appropriate for generators of local Lorentz invariance.
Also the spatial vierbein components \( {e^{A}}_{i} (i=1,2,3) \) must be associated
to conjugate momenta \( \pi ^{Ai}_e \) which either from an action
like (\ref{ActHP}) become constraints by themselves or must be expressible
in terms of other dynamical variables. On the other hand, there were
good reasons to retain the successful ADM formulation in terms of
lapse and shift. This program of {}``tetrad gravity'' started with
a seminal paper of Deser and Isham \cite{DEIS76} and has been
elaborated by several authors (exemplary references are \cite{HENE83}).
A common complication of approaches of this type of formulations 
has been the complexity
of the computations required for the determination of the Poisson
brackets.\footnote{
A shortcut proposed by Teitelboim \cite{TEIT73},
namely to rely on symmetry relations to avoid certain calculations,
turned out to be not applicable when terms of higher than linear order
in the constraints appeared \cite{CHHE88}.} 

To the best of our knowledge the first to consider tetrad gravity
in a first order formulation, e.g. with \emph{both} \( e_{A} \) and
\( \omega _{AB} \) as independent fields in the action (\ref{ActHP}),
were the authors of ref. \cite{CANI82}.
The condition of vanishing torsion appears as a second
class constraint which expresses the spin connection in terms of the
vierbeine and necessitates the introduction of Dirac brackets.

As mentioned already above, with the advent of Ashtekar's
variables \cite{ASHT86} the emphasis shifted towards the
connection components \( \omega _{ABI}\)  of $\omega_{AB} = 
\omega_{ABI}\, dx^I$. In the Palatini-type
action (\ref{ActHP}) the corresponding conjugate momenta 
\( \pi^{(AB)}_{\omega } \) become bi-vector\footnote{More on 
bi-vectors can be found in Appendix A.}
 components proportional to
$(e)^{AB} = e^{A}\wedge e^{B}$.  This can be taken care of by
adding the constraints, \( \phi ^{ij}=\epsilon _{ABCD}\pi
^{ABi}_{\omega }\pi ^{CDj}_{\omega }\approx 0 \), so that the
spatial vierbein variables are eliminated from the start (cf  
e.\ g.\  \cite{ASHT91}).  A similar elimination of the
vierbeine is the one of ref.  \cite{CDJ91} where only the
connection is kept as a variable already at the Lagrangian level.

The philosophy of our present paper is rather unconventional in the
sense that \textit{both} Cartan variables are treated 
{}``democratically'';
vanishing torsion does not appear by the Palatini mechanism, but
is imposed explicitly with the help of a Lagrangian multiplier 
which at first sight may appear as something of an overkill. 
However, it 
owes its basic idea to progress made recently in 1+1-dimensional dilaton
gravity theories, which includes spherically reduced gravity from
d dimensions, but also the string-inspired dilaton model of CGHS 
\cite{CGHS92}
and other simpler models, the most prominent being the one of Jackiw
and Teitelboim \cite{JACK85}. One of the present authors (W.K.) 
\cite{WKDS92} together with D.J.\ Schwarz 
at first in a special model with non-vanishing torsion \cite{KAVO90} 
realized 
the importance of using a temporal gauge in the Cartan variables which
led to extremely simple treatments of those models in the classical
and quantum case \cite{KULI97,REPO02}. In this
development the formulation as the linear derivative
action of a ``Poisson Sigma model'' \cite{SCST94,REPO02}
\begin{equation}
\label{Act2d}
S_{\left( LD\right) }=\int\limits _{\mathcal{M}_2}\,
 \left[ Xd\omega +X_{a}\, De^{a}\, +\epsilon \mathcal{V}\left( X^{a}X_{a,}X\right) \right] 
\end{equation}
 with real auxiliary fields $ X,X^{a}$ $( a=0,1;\, \, \,$ the
Hodge star of $ 2 d\omega$ becomes $R$, the Ricci
scalar in $ d=2 $; $ \epsilon =\frac{1}{2}\epsilon _{ab\,
}e^{a}\wedge e^{b} $ is the volume form) played an important
role.  Indeed all physically interesting models could be covered
by a {}``potential'' \( \mathcal{V} \) quadratic in \( X^{a} \), 
i.\ e.\ generically including nonvanishing torsion.  
By eliminating (algebraically) \( X_{a} \) and the
torsion dependent part of the spin connection \( \omega
_{ab}=\omega \epsilon _{ab} \) the usual torsionless dilaton
gravity action with dilaton field \( X \) could be reproduced,
the two formulations being locally and globally equivalent at
the classical, as well as at the quantum level
\cite{KULI97,KKL96}. However, on the basis of (\ref{Act2d}) 
even a background 
independent quantization of spherically reduced (torsionless 
Einstein) gravity has
been possible which actually reduces to (local) quantum
triviality in the absence of matter \cite{KULI97,REPO02}.
This formulation recently has turned out to be able to solve
as well some long-standing problems in $N=(1,1)$ dilaton supergravity
theories \cite{BEWK03}.
With respect to applications in string theory the first determination
of the full action (including fermionic fields) for $N=(2,2)$
supergravity also has been a consequence of this approach
\cite{BEWK04}.

Actually the unique mathematical properties of an action like (4) may 
be traced to the fact that it represents a (gravitational) 
example of a Poisson-Sigma model \cite{SCST94}. Such models share 
an important feature with Yang-Mills theory, namely that the 
algebra of Hamiltonian (secondary, first class) constraints closes without 
derivative terms on $\delta$-functions, a property which lies at 
the root of the relatively easy tractability of gravity theories 
in two dimensions. 

Thus it appears natural to generalize (\ref{Act2d}) from \(
\mathcal{M}_{2} \) to a four dimensional manifold \(
\mathcal{M}_{4} \) with 2-form auxiliary fields \( X^{AB},Y^{A}
\) replacing the zero-forms \( X,\, X^{a} \),  and, of course, a
polynomial in \( e^{A}\wedge e^{B} \) and \( X^{AB} \) instead
of the potential term $\epsilon\mathcal{V}$.  This generalization 
even may possess physical justification. Remembering that in the 
$1+1$ dimensional case one of the auxiliary fields ($X$ in 
(\ref{Act2d})) in spherically reduced GR owes its presence to the 
dilaton field, the same fact could be reflected in certain parts 
of the  new $X^{AB}$ which could be dilaton fields, related to 
some compactification mechanism from a gravity theory with 
dimension larger than four.

In the diploma thesis of one of the present authors
(H.S.) \cite{HSDI97} this idea passed a positive test: an action
of generic type (\ref{Act2d}) on \( \mathcal{M}_{4} \) with a
large number of free parameters was shown to lead --- up to a
topological term --- to Einstein (-deSitter) gravity, i.e.\ the
action \( S_{(HP)} \) of (\ref{ActHP}), when the fields \(
X^{AB},Y^{A} \) are eliminated by algebraic equations of motion
(e.o.m.).  In this preliminary work many questions were left
open: a systematic analysis of all possibly relevant topological
terms, a detailed comparison with previous formulations, which,
in principle, should be contained as special cases, a 
comparison of e.o.m.-s etc.  Therefore, a detailed confrontation
of a general Palatini-type action with the new formulation is the
subject of Section 2, resp.\ Section 3.

In the canonical analysis (Section 4) we have to treat a rather
large set of constraints, because, in contrast to $d=2$, only
\textit{part} of the components of \( X^{AB} \) and \( Y^{A} \)
become canonically conjugate momenta, the rest being Lagrange
multipliers, which turn out to be completely
determined.\footnote{This must be the case, of course, because
we do not introduce new physical degrees of freedom.} The
structural similarity to a gauge theory suggests a foliation
directly in our {}``time'' coordinate $x^0$ \textit{without} the
introduction of lapse and shift variables. This represents perhaps 
the most basic difference with respect to usual tetrad gravity, 
where almost everywhere\footnote{Among more recent exceptions we 
are only aware of ref.\ \cite{CHAR83}.} the ADM decomposition had been used. 
In our case  the resulting  
constraints are polynomial and their Poisson brackets (with one
{}``inessential'' exception) rather yield delta functions on the
right-hand side, just as for a non-abelian gauge theory and the 
aforementioned Poisson Sigma models --- and
not derivatives thereof as for the ADM constraints.  As we
proceed in this analysis we have opportunities to compare with
previous works and their relation to our present approach. 
Our analysis suggests that a linear combination of the initial 
constraints yields first class  ones 
that fulfil the Poincare algebra of special
relativity as algebra of the Poisson-brackets.

In the final Section 5 we summarize the results obtained, list
open problems and indicate some possible directions of future
work.  Some calculational details are relegated to the
Appendices: important formulas on vector forms are collected
in Appendix A, the rules of the {}``hat calculus'' are listed in Appendix
B.  Some complicated Poisson brackets are collected in
Appendix~C. All brackets have been computed by hand and checked 
by a computer algebra package. In Appendix D we give details on the determination of 
the Lagrange multipliers whose final results appear in the main 
text. 

%
%
%
\section{Alternative formulations of Einstein gravity}

In order to create a basis for the comparison of the actions in the
ensuing \mbox{e.o.m.-s} of our approach in terms of Cartan variables and
auxiliary fields (Section 3) we collect here some well-known formulations
in an appropriate notation.

First we note that $ S_{\left( HP\right) } $ of (\ref{ActHP})
can be written in several equivalent ways. Introducing the duality
 operation for the two-form components $F^{AB}$ of a 
bi-vector~$F$~\footnote{Appendices A and B should be consulted for more
details on bi-vectors, resp.\  the hat calculus.}
\begin{equation}
\label{DefHat}
\myHat{F_{AB}}:=\frac{1}{2}\,\epsilon _{ABCD}\, F^{CD},
\end{equation}
the identity 
\begin{equation}
\label{HatId}
\myHat{e^{AB}}=\Hodge{e}^{AB}\,,
\end{equation}
with $f^{AB}:=\left( f^{2}\right)^{AB}= (f\wedge f)^{AB}=
f^{A}\wedge f^{B}=-f^{BA} $,
 for an oriented orthonormal co-basis $e^A$ can be derived easily.
Equation (\ref{HatId}) allows to replace the hat by the Hodge star operation in
(\ref{ActHP}),
\begin{equation}
\label{ActHPHat}
S_{(HP)}=\int \myHat{e^{AB}}\wedge R_{AB}\,
=\int \Hodge{e^{AB}}\wedge R_{AB}\,.
\end{equation}
Hence, in terms of this type one can move not only the hat, but also the Hodge star operation freely
onto $ R_{AB} $, the anholonomic components of the Ricci tensor, 
 as well. It will be important in the next Section
that all such terms are equivalent.

Variation of $S = S_{(HP)} + S_{(M)}$, where $S_{(M)}$ is some
matter action, with respect to \( \delta e^{A} \)
yields\footnote{Here and thereafter we  neglect surface
terms.} the Einstein equations in the form 
\begin{equation}
\label{EOM_e}
\myHat{R_{BA}}\wedge e^{B}=\left( R_{AC}-\frac{1}{2}\,
R\, \eta _{AC}\right) \wedge \left( *e^{C}\right) 
=\frac{1}{2}T_{A}\; ,
\end{equation}
\noindent where the energy momentum 3-form \( T_{A} \) is defined
in terms of a matter action \( S_{(M)} \) as 
\begin{equation}
\label{VarActMatter}
\delta S_{(M)}=\int\limits _{\mathcal{M}_{4}}\left( \delta e_{A}\wedge T^{A}+
\frac{1}{2}\delta \omega _{AB}\wedge S^{AB}\right)\; .
\end{equation}
\noindent Variation of \( \delta \omega  \) leads to 
\begin{equation}
\label{EOM_omega}
\left( De^{2}\right) ^{AB}=\tau ^{A}\wedge e^{B}
-\tau ^{B}\wedge e^{A}=\frac{1}{2}S^{AB}\; ,
\end{equation}
\noindent with the torsion two-form (cf (\ref{DefTor}))
\begin{equation}
\label{DefTor2}
\tau ^{A}=de^{A}+\omega _{\, \, B}^{A}\,\wedge\, e^{B}.
\end{equation}
 For vanishing {}``spin current'' \( S^{AB} \) it can be shown
that the vanishing of the left hand side of (\ref{EOM_omega}) holds 
if and only if the torsion \( \tau ^{A} \) vanishes. 
This is the special case of a simple lemma (see Appendix A)  
which, in a less trivial application, also allows to determine the 
solution \( \tau ^{A} \) in terms of a 
non-vanishing \( S^{AB} \) in (\ref{EOM_omega}).

The first Bianchi identity \( ( \DKovA \tau )^A =: \DKovAK^{A}\,^{B}\, 
\tau_{B}=R^{A}\, ^{B}\wedge e_{B} \)
guarantees that adding an analogous {}``hat-less'' term to (\ref{ActHPHat})
with parameter \( \gamma \in \mathbb{R} \) 
\begin{equation}
\label{ActHPHatmod}
S'_{\left( HP\right) }=\int\limits _{\mathcal{M}_{4}}
\left( \myHat{e}^{\, AB}\wedge R_{AB} +\gamma e^{AB}\wedge 
R_{AB}\right)\; ,
\end{equation}
\noindent the e.o.m.-s (\ref{EOM_e}) are modified to
\begin{equation}
\label{EOM_emod}
\myHat{R_{AB}}\wedge e^{B}=\frac{1}{2}T_{A}- \gamma 
\left( \DKovA\tau \right) _{A}\; .
\end{equation}
\noindent The {}``torsion equation'' (\ref{EOM_omega}) also changes 
into 
\begin{equation}
\label{EOM_omegamod}
\left( \DKovA e^{2}\right) ^{AB}=
\frac{1}{2 \gamma ^{2} }\left( S^{AB}+\gamma 
\myHat{S^{AB}}\right)\; . 
\end{equation}
Thus, due to the first Bianchi identity also in that case the
Einstein equations are recovered for vanishing spin current \(
S^{AB} \) and vanishing torsion.  Identifying \( \gamma ^{-1} \)
with the Immirzi parameter \( \beta \) shows that Barbero's
Hamiltonian \cite{BARBERO} can be reproduced in this way.  Although
 (\ref{ActHPHat}) and (\ref{ActHPHatmod}) yield the
same classical solutions to the \EOMs,{} at the quantum level, of course,  
the theories would be expected to be different.\footnote{We use 
this cautious formulation because according to the work of refs.\ 
\cite{Alexandrov} and \cite{Alex+Vas} there exists an 
Ashtekar-like formulation where the connection is the one of the 
full SO(3,1) and where $\gamma$ appears as an ``anomaly 
candidate'', to be removed by renormalization.}

Further terms which could be added to $S_{(HP)}$ without changing
the e.o.m.-s are two topological ones: the Gauss-Bonnet-Chern form
\begin{equation}
\label{DefGaussBonnet}
L _{\left( GBC\right) }=\frac{1}{16\pi ^{2}}\myHat{R_{AB}}\wedge R^{AB}
\end{equation}
\noindent and the Pontrijagin form 
\begin{equation}
\label{DefPontrijagin}
L _{\left( P\right) }=\frac{1}{8\pi ^{2}}R_{AB}\wedge R^{AB}.
\end{equation}
By analyzing all terms quadratic in $ R_{AB} $ according to the
rules of Appendix B  
(with hats and stars distributed over both factors)
it can be verified by means of the second Biancchi identity 
$ \left( \DKovA R\right) ^{AB} = (\myHat{\DKovA R^{AB}})=0 $
that (\ref{DefGaussBonnet}) and (\ref{DefPontrijagin}) are the only 
possibilities. Thus
the most general Palatini-type action to be compared with our own
formulation below can be written as 
\begin{eqnarray}
\label{ActHPext} 
S^{\left( ext\right) }_{(HP)} & =  \int\limits_{\mathcal{M}_{4}}
 \Big( & R_{AB}\wedge \myHat{e^{AB}}+ 
 \gamma R_{AB}\wedge e^{AB}+\rho R_{AB}\wedge R^{AB}+  \nonumber \\
 &  &  \sigma R_{AB}\wedge \myHat{R^{AB}}+\frac{\Lambda }{12}
 \myHat{e^{AB}}\wedge e_{AB}\Big).
\end{eqnarray}

\noindent with general parameters \( \gamma ,\rho ,\sigma \).
The last term with the
cosmological constant \( \Lambda \) allows to include the
Einstein-deSitter case as well.

For completeness we also place the complex formulation of real
GR \cite{ASHT86} into the present context.  For the (anti-) selfdual connections
\begin{equation}
\label{2.14}
\omega ^{\pm }_{AB}:=\frac{1}{2}\left( \omega _{AB}\pm i\myHat{\omega _{AB}}\right) 
\end{equation}
\noindent only three of the six components of \( \omega ^{\pm }_{AB} \)
are linearly independent over $\Complex $. Analogously, the 
(anti-)selfdual curvature is defined as:
\begin{equation}
\label{2.15}
R^{\pm }_{AB}\left( \omega \right) :=
\frac{1}{2}\left( R_{AB}\left( \omega \right) \pm i\myHat{R_{AB}}
\left( \omega \right) \right) \; .
\end{equation}
Applying the hat operation (\ref{DefHat}) to (\ref{2.15}) the HP action
(\ref{ActHP}) can be rewritten as 
\begin{equation}
\label{2.16}
S_{(HP)}=i\int\limits _{\mathcal{M}_{4}}\left[ -R_{AB}^{+}+R_{AB}^{-}\right] 
\wedge e^{AB}.
\end{equation}
Choosing \( \omega ^{\pm }_{0a} \) as a basis (\( A= 0,a  \)
etc.). from (\ref{2.14}) and (\ref{2.15}) the components of the 
(anti-)selfdual curvature  can be expressed as
\begin{equation}
\label{2.17}
R_{0a}^{\pm }=d\omega ^{\pm }_{0a}\, \pm i\epsilon _{0a}\, ^{cd}\omega ^{\pm }_{0c}\wedge \omega _{0d}^{\, \, \, \pm },
\end{equation}
\noindent with the structure constants \( \epsilon \DO{0a}\UP{cd} \)
implying that the Lie group for \( \omega ^{\pm }_{0a} \) 
is \( SO\left( 3,\mathbb {C}\right)  \).
Ashtekar's variables arise if one of the two terms in the real action
(\ref{2.16}) is dropped so that the contribution of the 
variation with respect to the
{}``complex conjugate'' connection one-form disappears. In 
contrast ours will be a full SO(3,1) gauge formulation as shown 
below.

\section{Actions with auxiliary fields}

\subsection{Equivalent Action}

Motivated by the success of the analogous $1+1$ dimensional dilaton theory (\ref{Act2d})
the basic ansatz of our approach 
\begin{equation}
\label{ActLD}
S_{\left( LD\right) }= S_{\left( X\right) }+S_{\left( Y\right) }
+S_{\left(\tilde{\Lambda}\right) }\,,
\end{equation}
with
\begin{eqnarray}
S_{\left( X\right) }\left( X,\omega ,e\right) 
=\int\limits _{\mathcal{M}}\Bruch{1}{2}\left( a_{0}X^{AB}\wedge X_{AB}
+a_{1}X^{AB}\wedge \Hodge{X_{AB}}+
 a_{2}X^{AB}\wedge \myHat{X_{AB}}+\right.  &  & \nonumber \\
a_{3}X^{AB}\wedge \Hodge{\myHat{X_{AB}}}
+b_{0} X^{AB}\wedge R_{AB}+ b_{1}X^{AB}\wedge \Hodge{R_{AB}} \, + \,\,
 & \nonumber \\
\left. b_{2}X^{AB}\wedge \myHat{R_{AB}}+b_{3}X^{AB}\wedge \Hodge{\myHat{R_{AB}}}+c_{0}X^{AB}\wedge e_{AB}+
c_{2}X^{AB}\wedge \myHat{e_{AB}}\right) , &  & \label{ActLDSX} 
\end{eqnarray}

\begin{equation}
\label{ActLDSY}
S_{\left( Y\right) }=\int\limits _{\mathcal{M}} 
Y^{A}\wedge \left( \DKovA e \right) _{A}\,, 
\end{equation}

\begin{equation}
\label{ActLDSCC}
S_{\left(\tilde{\Lambda} \right) }=\Bruch{\widetilde{\Lambda }}{12}\,
\int\limits _{\mathcal{M}}
 \myHat{e^{AB}}\wedge e_{AB}\,,
\end{equation}
\noindent consists of at most quadratic expressions in auxiliary two-form
fields \( X^{AB},\, Y^{A}, \) the curvature \( R^{AB} \) and the
bi-vector \( e^{AB} \). Eq. (\ref{ActLDSY}) explicitly imposes the condition
of vanishing torsion, although, as is well-known (cf also Section
2), the Palatini mechanism implies this anyhow. However, the advantage is
that in this way, at least at the start, an independent momentum variable,
canonical conjugate to $ e_{Ai} $ has been introduced. The eventual
appearance of a cosmological term has been foreseen by the inclusion
of $ S_{\left(\tilde{\Lambda} \right) } $ in (\ref{ActLDSCC}). It should be
noted, however, that the \oquote effective\equote\ cosmological constant 
$\Lambda $ in
the equivalent Einstein-deSitter theory will also acquire contributions
from combinations of the constants 
$ a_{i},b_{i}\left( i=1,...4\right) ,\, c_{0},c_{2} $.
The restriction to coefficients $ c_{0} $ and $ c_{2} $ only
in the last two terms of (\ref{ActLDSX}) originates from the identity
(\ref{HatId}).
 In the rest of (\ref{ActLDSX}) simply all independent starry and hatted 
terms have been collected.  It should be noted that 
the formulations of gravity as
a BF-theory exhibit a certain similarity to our ansatz, the
main difference being that our two-form auxiliary fields
$X\Up{AB}$ are not simply proportional to $e\UP{AB}$ (cf eq.\  
(\ref{EOM_solX}) below)  \cite{PLEB77} and $Y^A$ (at least 
initially) does not vanish. 
A more compact form of (\ref{ActLDSX}) is obtained by noting
that the hat as well as the Hodge star operator --- in our
applications to bi-vectors only --- act like (mutually
commuting) imaginary units of two independent complex structures
\( (\CA {i}^{2}:=-1, \CA {j}^{2}:=-1, \CA{i j} = \CA{ji}, \left(
a+\CA {i}b\right) ^{*}:=a-\CA {i}b,\left( c+\CA {j}d\right)
^{\wedge }:=c-\CA {j}d) \)
\begin{eqnarray}
a\Hodge{X_{AB}} & =: & a\CA{i}X_{AB},\nonumber \\
a\, \myHat{X_{AB}} & =: & a\CA {j}X_{AB},\label{3.5}  \\
a\Hodge{\myHat{X_{AB}}} & =: & a\CA{ij}X_{AB.} \nonumber
\end{eqnarray}
\noindent This permits abbreviations of the form 
\begin{equation}
\label{Defa}
\Va = a_{0}+a_{1}\CA {i}+\CA {j} \left( a_{2}+ \CA {i}a_{3}\right)=:
 \CAi{a}  + \CA{j} \CAii{a} 
\end{equation}
\noindent in (\ref{ActLDSX}) which may be considered as elements
(\ref{Defa})  of a
{}``bi-complex'' algebra $\mathbb{C}^2$:
with non-trivial, \ie\ different from zero, not necessarily invertible 
elements $\Va, \Vb, \Vc$ (for a first analysis cf \cite{HSDI97}):
\begin{equation}
\label{3.7}
S_{\left( X\right) }=\int \Bruch{1}{2} \left( 
\left( \CA{a}X\right) ^{AB}\wedge X_{AB}+
\left( \CA {b}X\right) ^{AB}\wedge R_{AB} + 
\left( \CA {c}X \right) ^{AB}\wedge e_{AB}\right)
\end{equation}
\noindent The resulting \EOMs\ from variation of \( \delta X_{AB} \)
\begin{equation}
\label{EOM_solX}
X^{AB}=-\frac{\CA {a}}{2}^{-1}\left( \CA {b}R^{AB}+\CA {c}e^{AB}\right) 
\end{equation}
\noindent are solved with the 
inverse of (\ref{Defa}),
\begin{equation}
\label{3.9}
\CA {a}^{-1}= 
\frac{\CAi{a}-\CA{j}\CAii{a}}{{\CAi{a}}\UP{2}+{\CAii{a}}\UP{2}}\,
=\,|a|^{-4} \left( {\CAi{a}}\UP{2}+ {\CAii{a}}\UP{2}\right)
^{*}\CA {a}^{\wedge }\; ,
\end{equation}
\noindent where
\begin{eqnarray}
\CAi{a} & := & a_{0}+ \CA {i} a_{1},\nonumber \\
\CAii{a} & := & a_{2}+\CA {i} a_{3},\label{3.10} \\
|a|^4  & := & 
\left( {\CAi{a}}\UP{2}+ {\CAii{a}}\UP{2}\right)
\left( {\CAi{a}}\UP{2}+ {\CAii{a}}\UP{2}\right) \Up{*} \; \geq\; 
0.\nonumber
\end{eqnarray}
The existence of (\ref{3.9}) and hence of 
(\ref{EOM_solX}) is guaranteed if the $a_{i}$'s 
are chosen such that \( |a| \neq 0 \). 

The linearity of (\ref{EOM_solX}) 
guarantees  that the variational principle remains unchanged by 
the insertion of (\ref{EOM_solX}) in (\ref{3.7}). Thus the 
 extended HP-action (\ref{ActHPext}) is reproduced if after insertion
of (\ref{EOM_solX}) the coefficients in
\begin{eqnarray}
S_{\left( X\right) } & =   \int\limits _{\mathcal{M}}& 
\Bruch{1}{2}\left[ 
 \left( \CA {k}R_{AB}\right) \wedge R^{AB}+\left( \CA {l}R_{AB}\right) \wedge e^{AB}+\right. \nonumber \\
 &  & \,\,\left. \left( \CA {m}e_{AB}\right) \wedge e^{AB}\right] \label{3.11} 
\end{eqnarray}
\noindent are chosen as
\begin{eqnarray}
\CA {k} & = & -\frac{1}{4}\CA {a}^{-1}\CA {b}^{2}= 
2\,(\rho +\CA {j}\sigma\,)\, ,\nonumber \\
\CA {l} & = & -\frac{1}{2}\CA {a}^{-1}\CA {b}\CA {c}=
2\,(\gamma +\CA{j})\;,\label{3.12} \\
\CA {m} & = & -\frac{\CA {a}^{-1}}{4}\CA {c}^{2}=m_{0}+\CA 
{j}m_{2}\, .
\label{3.13}
\end{eqnarray}
In the ans\"{a}tze (\ref{3.12}) for the r.h.s.\  
expressions involving the second ``imaginary'' unit 
$\CA{i}$ do not appear. The reason is that there are no topological terms of 
the curvature 2-form $R^{AB}$ containing the Hodge star in 
(\ref{ActHPext}). This 
implies that the bi-complex algebra $\mathbb{C}^2$ may be 
effectively projected upon $\mathbb{C}$ (cf the property of the full 
solution, (\ref{3.17}) below).

\noindent With the ansatz (\ref{3.13}) for $\CA {m}$ the identity 
\begin{equation}
\label{3.14}
4\CA {km}=\CA {l}^{2}
\end{equation}
\noindent directly determines 
\begin{equation}
\label{3.15}
m_{2}= c_0 + c_2\, \gamma = 
\frac{2\rho \gamma +\sigma \left( 1-\gamma ^{2}\right) }
{2\left( \sigma ^{2}+\rho ^{2}\right) }.
\end{equation}
In (\ref{3.13}) \( m_{0} \) is irrelevant because 
\( e_{AB}\wedge e^{AB}\equiv 0 \) , 
whereas (\ref{3.14}) together with (\ref{ActLDSCC}) provide 
contributions to the cosmological constant \( \Lambda  \) in (\ref{ActHPext}):
\begin{equation}
\label{3.16}
\Lambda = \widetilde{\Lambda } + 6 m_{2}
\end{equation}
The identity (\ref{3.14}) implies that the coefficients 
\( \CA {a},\CA {b},\CA {c} \)
in the action (\ref{3.7}) cannot be obtained uniquely from (\ref{3.12})
and (\ref{3.13}). Inserting the full solution with 
\( \CA {q}\in \mathbb{C}^2 \)
\begin{equation}
\label{3.17}
\CA {a}=-4\CA {q}^{2}\CA {k},\, \, \CA {b}=-4\CA {q}\CA {k},\,\, 
\CA {c}=-2\CA {q}\CA {l}
\end{equation}
\noindent into the action \( S_{(X)} \) of (\ref{ActLDSX}) shows that
the arbitrary invertible element \( \CA {q} \) corresponds to a 
rescaling    
\( \CA {q}X^{AB}\rightarrow X^{AB} \) which is always 
possible\footnote{This possibility removes the only variable 
which could be an element of the whole bi-complex algebra. 
Nevertheless, the latter is useful to keep the discussion very 
simple but still completely general, at the intermediate steps 
leading to (\ref{3.18}), (\ref{3.19}), but also in the 
Hamiltonian analysis below.}.
If the coefficient $\CA{b}$ of the term $ X_{AB}\wedge R^{AB} $ is chosen to
be $1$ as in the $2d$ counterpart (\ref{Act2d}) we may fix the overall
factor as \( -4\CA {kq}= \CA{b} = 1 \). Therefore, in
\begin{equation}
\label{3.18}
S_{\left( X\right) }=\int\limits _{\mathcal{M}}
\Bruch{1}{2}\,X_{AB}\wedge\left( R^{AB}+\CA {a}X^{AB}+\CA {c}e^{AB}\right) 
\end{equation}
\noindent the coefficients
\begin{equation}
\label{3.19}
\CA {a}=-\frac{\rho -\CA {j}\sigma }{8 \left( \rho ^{2}+\sigma ^{2}\right) },\, \,
 \CA {c}=
\frac{\rho \gamma +\sigma +\CA {j}\left( \rho -\sigma \gamma \right) }
{2\left( \rho ^{2}+\sigma ^{2}\right) }
\end{equation}
\noindent together with (\ref{3.15}) provide agreement with the
generalized Palatini-type action (\ref{ActHPext}).  
As at least one
of the coefficients $ \rho $ or $\sigma $ must be nonvanishing
the appearance of at least one topological term
(\ref{DefGaussBonnet}) or (\ref{DefPontrijagin}) is mandatory. 
The coefficient $\CA {c}$ cannot be made to disappear altogether. 
Not surprisingly, also the terms quadratic in \( X \) must be present
always.  A solution with $\CA{b}=0$ is not possible (cf 
(\ref{3.12})). It would correspond to the BF-model of \cite{PLEB77}. 
Our present formulation  (\ref{3.18}) can be 
interpreted as the most comprehensive version of a generalized 
BF-model \cite{SMOL63}.
On the other hand, 
dropping the term with the Immirzi parameter ($\gamma = 0$) in 
(\ref{ActHPHatmod}) according to (\ref{3.19}) is perfectly 
consistent with the ansatz (\ref{ActLD}), resp.\ (\ref{3.7}). 

An alternative nice parametrization of (\ref{3.19}) follows from
\( \rho =r \, \cos\, \varphi ,\, \sigma =r \, \sin\, \varphi  \)
\begin{equation}
\label{3.20}
\CA {a}=-(8r)^{-1}e^{-\CA {j}\varphi },\, \, \CA {c}=\left( 2r
\right) ^{-1}\, e^{-\CA{j}\varphi }\, ( \gamma+\CA {j}\,)
\end{equation}
\noindent which interpolates by variation of $\varphi$ between vanishing 
\( \rho , \) resp.\ \( \sigma  \). It also shows that the most 
``symmetric'' solution for \( \CA{a} \) and \( \CA{c} \) is the 
one for $\gamma = 0$. 

\subsection{Equations of motion from auxiliary field action}

The algebraic equivalence of the actions (\ref{3.18}) and (\ref{ActHPext}) 
--- even by only linear relations --- guarantees identical
dynamical content. Nevertheless, the explicit form of the \EOMs\
from  (\ref{3.18}) will be very useful 
 for comparison with the ones derived from the Hamiltonian.
Variations of $ \delta Y_{A},\, \delta X_{AB},\, \delta \omega _{AB} $
and $ \delta e_{A} $, respectively yield 
\begin{equation}
\label{3.21}
\tau ^{A}=0,
\end{equation}
\begin{equation}
\label{3.22}
2\CA {a}X^{AB}+R^{AB}+\CA {c}e^{AB}=0,
\end{equation}
\begin{equation}
\label{3.23}
\left( DX\right) ^{AB}\,+\,\left( Y^{A}\wedge e^{B}-Y^{B}\wedge e^{A}\right) =0,
\end{equation}
\begin{equation}
\label{3.24}
\left( DY\right) ^{A}\,+\,\CA {c}X^{AB}\wedge e_{B}+\frac{\widetilde{\Lambda }}{3!}\epsilon \UP{A}_{BCD}e^{B}\wedge e^{CD}=0,
\end{equation}
\noindent which allow further simplification. The commutativity of
the hat operation with the covariant derivative applied to (\ref{3.22})
leads to\begin{equation}
\label{3.25}
2\CA {a}\left( DX\right) ^{AB}+\left( DR\right) ^{AB}+\CA {c}\left( De^{2}\right) ^{AB}=0,
\end{equation}
\noindent where the second term vanishes by Bianchi's second identity.
The third one is eliminated by (\ref{EOM_omega}) and (\ref{3.21}). 
In (\ref{3.23}) this leads to the disappearance of the first term, and 
the rest by the same argument as for (\ref{EOM_omega}) 
\( \left( S^{AB}=0\Leftrightarrow \tau ^{A}=0\right)  \) forces \( Y^{A}=0. \)

Therefore, the last two equations (\ref{3.23}) and (\ref{3.24})
may be replaced by
\begin{equation}
\label{YsNull}
Y^{A}=0\; ,
\end{equation}
\begin{equation}
\label{3.27}
\CA {c}X^{AB}\wedge e_{B}+\frac{\widetilde{\Lambda }}{6}
\epsilon \UP{A}_{BCD}e^{B}\wedge e^{CD}=0.
\end{equation}
It is straightforward to verify that the set (\ref{3.16}), 
(\ref{3.21}), (\ref{3.22}), (\ref{YsNull}), (\ref{3.27}) is equivalent  
 to the \EOMs\ from (\ref{ActHPext}). We also emphasize that the 
 vanishing of torsion by (\ref{3.21}) as well as of the related 
 auxiliary field $Y^A$, according to (\ref{YsNull}), in our 
 approach occurs  on-shell only. 

%
%
\section{Hamiltonian Analysis}

\subsection{Canonical Hamiltonian}

The action (\ref{3.18}) is of  Hamiltonian form. Due to the
linearity in the derivatives the momenta are directly associated to
(part of the components of) the auxiliary fields \( X^{AB},\, Y^{A}. \)
The identification in such cases is an allowed short cut 
(cf \cite{HSDI97}), although,
 strictly speaking the corresponding relations are second class constraints.

Using the transcription rule  in the typical two-form multiplications
\begin{equation}
\label{4.1}
U\wedge V=\left( U_{0i}\, V_{jk}+U_{jk}V_{0i}\right) 
\frac{\epsilon^{ijk}}{2}d^{4}x
\end{equation}
\noindent for $\epsilon ^{0ijk}=-\epsilon ^{ijk}$ (the Levi-Civitá symbol 
with $\epsilon^{123}=1$) to obtain the different pieces 
of the action (\ref{3.18}) in terms of corresponding parts of the
Lagrangian density $\quad$
$ \mathcal{L}=\mathcal{L}_{\left( X\right) }+
 \mathcal{L}_{\left( Y\right) }
 +\mathcal{L}_{\left(\tilde{\Lambda} \right) } $\, ,
the result is
\begin{eqnarray}
\mathcal {L}_{\left( X\right) } & = 
 \Bruch{\epsilon ^{ijk}}{4}\big\{ &
   X\UP{AB}_{0i}
   \left( 2\CA {a}X_{ABjk}+R_{ABjk}+\CA {c}e_{ABjk}\right) +
  \nonumber \\
 &  & 
  X\UP{AB}_{jk}\left( \partial\Do{0} \omega _{ABi}\right) +
      e_{AB0i}\CA {c}X\UP{AB}_{jk} +  \label{4.2}
 \\
 &  & 
 \omega _{AB0}
  \left( \partial _{i}X\UP{AB}_{jk} + 
  \omega ^{A}_{\, \, Ci} X\UP{CB}_{jk} - \omega\UP{B}_{Ci}
         X\UP{CA}_{jk}\right) \big\}\,, 
\nonumber\\
\nonumber\\
\mathcal{L}_{\left( Y\right) } & =  
\Bruch{\epsilon ^{ijk}}{2}\big\{&
  Y\UP{A}_{0i} \tau _{Ajk} + 
  \left( \partial_{0} e_{Ai} \right) Y\UP{A}_{jk} +  \nonumber \\
 &  & 
   e_{A0} \left( \partial_{i}\, Y\UP{A}_{jk} +
           \omega \UP{A}_{Ci}\, Y\UP{C}_{jk} \right) +
 \label{4.3} 
\\
 &  & 
  \Bruch{1}{2}\omega _{AB0}
   \left( Y\UP{A}_{jk}e_{\, i}^{B}-Y\UP{B}_{jk}\, e\UP{A}\Do{i}\right) 
  \big\} ,
\nonumber\\
\nonumber\\
\label{4.4}
\mathcal {L}_{\left(\tilde{\Lambda}\right)  } & = 
  \, \,
& \epsilon ^{ijk}\Bruch{\widetilde{\Lambda }}{6}\,
  e_{A0}\epsilon \UP{A}_{BCD} \,e\UP{B}_{i}\,e^{C}_{j}\,e_{k.}^{D}
\end{eqnarray}
In eqs.\ (\ref{4.2}) and (\ref{4.3}) as in all previous 
computations total divergencies again have been disregarded. 

The momenta conjugate to \( \xo_{ABi}, \) resp.\ \( \xe_{Ai} \),
with \( \Ko^{0} \) defined as a time coordinate for any coordinate map 
$\left(\Ko\Up{I}\right)$, can be read off from (\ref{4.2}) and (\ref{4.3}) 
\begin{equation}
\label{4.5}
\pi^{ABi}_\omega = 
\Pi^{ABi} := \HalbF{\epsilon ^{ijk}} X\UP{AB}\Do{jk} \, ,
\; A<B\; ,
\end{equation}
\begin{equation}
\label{4.6} 
\pi^{Ai}_e =\pi\Up{Ai}:=\HalbF{\epsilon ^{ijk}} Y\UP{A}_{jk}\;,
\end{equation}
whereas the conjugate momenta of 
\( \xo _{AB0} \) 
and
\( \xe_{A0} \) vanish as primary constraints. 
Therefore, we may effectively drop the set $e_{A0}, \omega_{AB0},
 X_{AB0i}, Y_{A0i}$ from the ranks of canonical variables, 
 but treat them as Lagrange multipliers.
This reduces the dimension of our phase space
considerably from $\textrm{N}\Do{ph}= 2\cdot(16+24+36+24)=200$ 
to $\textrm{N}\Do{ph}= 2\mal(12+18)=60$.
 The canonical Hamiltonian (from now on we use: 
\(
\partial _{0}f=: \dot{f},\, x_{ABi}:=X_{AB0i},\, y_{Ai}:=Y_{A0i}    
\) 
)
\begin{eqnarray}
H_{(can)} & = & \int d^{3}x \left( \Halb \Po \Up{ABi}\,\dot{\xo}_{ABi}+ 
\Pe ^{Ai}\,\dot{e}_{Ai}-\mathcal{L}\right) = \nonumber \\
 & = & \int d^{3}x\left( \xe_{A0}\eC\Up{A} + 
\Halb \xo_{AB0} \oC\Up{AB} + 
\Halb x_{ABi}\XC\Up{ABi} + y_{Ai}\YC\Up{Ai} 
\right)
\label{Hcan} 
\end{eqnarray}
depends on the constraints
\begin{equation}
\label{4.8}
\oC ^{AB}=\oCX\Up{AB}+\oCY\Up{AB} \approx 0 \;,
\end{equation}
\begin{equation}
\label{4.9}
\eC^{A}=\eCX\Up{A}+\eCY\Up{A}+ E^A_{(\tilde{\Lambda})} 
\approx 0 \; ,
\end{equation}
where in all cases the origin of the respective
contributions
\begin{eqnarray}
\label{DefoCX}
-\oCX\Up{AB} & = & \DKovAK_i \Po\Up{ABi} = \partial _{i}\Po\Up{ABi} + 
\xo\UP{A}_{Ci} \Po\Up{CBi} - \xo\UP{B}_{Ci}\Po^{CAi}\,,\\
\label{DefoCY}
-\oCY\Up{AB} & = & \Pe^{Ai} \xe\UP{B}_i- \Pe^{Bi} \xe\UP{A}_i \,,\\
-E_{\left( X\right) }^{A} & = & \CA{c} \Po^{ABi} \xe_{Bi} \,,\\
-E_{\left( Y\right) }^{A} & = & \DKovAK_{i} \Pe^{Ai} = \partial_{i}\Pe^{Ai} + \xo\UP{A}_{Bi} \Pe^{Bi}\,,\\
-E_{\left(\tilde{\Lambda} \right) }^{A} & = & \Bruch{\widetilde{\Lambda }}{6} 
\epsilon\UP{ABCD} \epsilon^{ijk} \xe_{Bi}\xe_{Cj}
\xe_{Dk} \label{4.14} 
\end{eqnarray}
is indicated, and 
\begin{equation}
\label{4.15}
-\YC^{Ai} = - \YC^{Ai}_{(Y)} = \Halb\Tor\UP{A}_{jk}\,\epsilon ^{ijk} = 
\left( \partial _{j}\xe\Up{A}_{k}+\xo\UP{A}_{Bj}\xe^{B}_{k}\right)
 \epsilon ^{ijk} \approx 0 \; ,
\end{equation}
\begin{equation}
\label{DefXC}
-\XC^{ABi} = - \XC_{(X)}^{ABi} = 
2\,\CA{a} \Po\Up{ABi}+ R^{ABi}+ \CA {c}\xe^{ABi} \approx 0\;.
\end{equation}
We made use of the convenient abbreviation (already suggested by 
(\ref{4.5}))
\begin{equation}\label{RaumindexHoch}
W\UP{ABi}:= \Bruch{\epsI}{2} \, W\UP{AB}_{jk} 
\end{equation}
for any two-form $W\UP{AB}$. With that definition we have e.g.\ 
from (\ref{4.5}), (\ref{4.6}) 
\begin{equation}\label{RaumindexHochBsp}
  X\UP{ABi}=\Po\UP{ABi}, \quad 
  Y\UP{Ai}=\Pe\UP{Ai} \,.
\end{equation}
By counting the degrees of freedom\footnote{Our construction of the 
theory ensures that we have two  
degrees of freedom as ordinary Einstein-de Sitter GR (see Section 
4.3 below).} one immediately  infers that these $6+4+12+18=40$
 constraints cannot all be independent first class for the 
$30$ variables $(\xe_{Ai}, \xo_{ABi})$, and thus their Poisson bracket 
algebra will not close.

The Poisson brackets\footnote{The appropriate formulas of Appendix C 
may be consulted.} of
the six constraints $\oC\Up{AB}$, the constraints conjugate to
$\xo_{AB0}$, with all secondary constraints become
(e.g.~$E^{A'}$ is a shorthand for $E^{A'}(x^0,x^{i'}$), 
$\dfs:=\delta^3 (x^i-x^{i'}$)
\begin{align}
\PK{\oC\UP{AB}, \eC\UP{A'}} &= (\eta\UP{AA'} \eC\UP{B}-\eta\UP{BA'} 
\eC\UP{A})\,\dfs\,, 
\label{PKoe}\\
\PK{\oC\UP{AB}, \oC\UP{A'B'}} &= (\eta\UP{AA'} \oC\UP{BB'}-\eta\UP{BA'} \oC\UP{AB'} +
\eta\UP{BB'} \oC\UP{AA'}-\eta\UP{AB'} \oC\UP{BA'})\,\dfs\,,  \label{PKoo}\\
\!\!\!\!\PK{\oC\UP{AB}, \XC\UP{A'B'i'}} & =  
  (\eta\UP{AA'} \XC\UP{BB'i'}\!-\!\eta\UP{BA'} \XC\UP{AB'i'} \!+\!
   \eta\UP{BB'} \XC\UP{AA'i'}\!-\!\eta\UP{AB'} \XC\UP{BA'i'})\,\dfs\,, 
\label{PKoX}\\
\PK{\oC\UP{AB}, \YC\UP{A'i'}} &=
 (\eta\UP{AA'} \YC\UP{Bi'}-\eta\UP{BA'} \YC\UP{Ai'})\,\dfs\,,
\label{PKoY} 
\end{align}
and whence are weakly equal to zero. Eqs.\ 
(\ref{PKoe})--(\ref{PKoY}) suggest that 
$\oC\UP{AB}$ are proper generators of local Lorentz transformations.

The computation of the Poisson brackets of the $34$ remaining, 
non-first class secondary constraints, on the one hand, yields
\begin{align}
\PK{\XC\UP{ABi}, \XC\Up{A'B'i'}}&=0\,,\label{PKXX}\\
\PK{\YC\UP{Ai}, \YC\UP{A'i'}}&=0\,,\label{PKYY}\\
\PK{\eC\UP{A}, \eC\UP{A'}} &= - 
 (\PHNum{c} \XC\UP{AA'} + \Lambda \eps\UP{AA'}\Do{CD} 
 \YC\UP{Ci}\xe\UP{D}_i)\,\dfs \,, \label{PKee}
\end{align}
which also weakly vanish.  But, on the other hand, 
upon conservation of the secondary constraints 
$\eC\UP{A}, \oC\UP{AB}, \XC\UP{ABi}, \YC\UP{ABi}$ the remaining 
brackets 
\begin{align}
\!\!\!\!\PK{\XC\UP{ABi}, \YC\UP{A'i'}} =& 
 2\PHNum{a}(\eta\Up{AA'}\xe\UP{B}\Do{k'} - \eta\Up{BA'}\xe\UP{A}\Do{k'} ) 
 \epsilon^{i'i\,k'} \dfs\,,\label{PKXY}\\
\!\!\!\!\PK{\eC\UP{A}, \YC\Up{A'i'}}=&
 (\XC\Up{AA'i'}+2\PHNum{a}\,\Po\Up{AA'i'}+3c_2\PHNum{j}\,\xe\Up{AA'i'})\,\dfs\,,
\label{PKeY}\\
\!\!\!\!\!\PK{\eC\UP{A}, \XC\Up{A'B'i'}}=&
 \big(
 \PHNum{c}(\eta\Up{AA'}\YC\Up{B'i'}-\eta\Up{B'A}\YC\Up{A'i'})+ \nonumber \\
 & \,\,2\PHNum{a}(\eta\Up{AA'}\Pe\Up{B'i'}-\eta\Up{B'A}\Pe\Up{A'i'})
 \big)\,\dfs\,,
\label{PKeX}
\end{align}
imply the appearence of ternary constraints and the fact that
some of the secondary ones are second class.  It should
be noted that in two-dimensional gravity (\ref{Act2d})
\cite{REPO02} the Poisson bracket algebra of the secondary
constraints yields first class ones only, i.e.\ brackets 
like (\ref{PKXY}) -- (\ref{PKeX}) do not appear there.

\subsection{Ternary constraints}
As a consequence of the Poisson brackets (\ref{PKeY})
and (\ref{PKeX}) 
the time evolution of the constraints $\eC\UP{A}$ 
is given by
\begin{equation}\label{eTimeEvolution}
\begin{split}
\dot{\eC}\UP{A}\,\approx\,& y_{A'i'} 
(2\Va\,\Po\UP{AA'i'}+3c_2\PHNum{j}\,\xe\Up{AA'i'}) 
+\, x_{A'B'i'}\Va\,(\eta\UP{AA'} \Pe\UP{B'i'} - \eta\UP{AB'} \Pe\UP{A'i'})\,
 \\[1mm] \sg\,&
2\Va\,(y_{A'i'} \Po\UP{AA'i'} + x\UP{A}_{B'i'}\,\Pe\UP{B'i'} ) 
+ 3c_2\,\PHNum{j} y_{A'i'}\,\xe\Up{AA'i'} \,\sg\,0 \,, 
\end{split}
\end{equation}
and the time evolution of $\XC\UP{ABi}$ follows from (\ref{PKeX}) and
(\ref{PKXY}) as  
\begin{equation}\label{XTimeEvolution}
\begin{split}
\dot{\XC}\UP{ABi}\,\approx\,& y_{A'i'} 2\Va\,
(\eta\UP{AA'}\,e\UP{B}_{k'}-\eta\UP{BA'}\,e\UP{A}_{k'})\eps\UP{ii'k'}\,+\,
e_{A'0}\,2\Va\,(\eta\UP{AA'} \Pe\UP{Bi} - \eta\UP{BA'} \Pe\UP{A'i} )
\\[1mm]
\,\sg\,& 2\Va \left( \eps\UP{ii'k'}
 (y\UP{A}_{i'} \,e\UP{B}_{k'}-y\UP{B}_{i'} \,e\UP{A}_{k'}) \,+
\,
e\UP{A}_{0}\Pe\UP{Bi}-e\UP{B}_{0}\Pe\UP{Ai}\right)\,\sg\,0\,.
\end{split}
\end{equation}
Finally the evolution equation for the twelve $\YC\UP{Ai}$ are found
from  (\ref{PKXY}) and (\ref{PKeY}) 
\begin{equation}
\label{YTimeEvolution}
\begin{split}
\dot{\YC}\UP{Ai}\,\approx\,& 
x_{A'B'i'}\,\Va\,(\eta\UP{AA'}\,e\UP{B'}_{k'}
-\eta\UP{AB'}\,e\UP{A'}_{k'})\eps\UP{ii'k'}\,+
(2\Va\,\Po\UP{AA'i}+3c_2\,\Vj\,e\UP{AA'i'})\,e_{A'0}\,
\\[1mm] 
\,\sg\,&  2\Va\, (x\UP{AB'}_{i'}\,e_{B'k'}\,\eps\UP{ii'k'} + 
\Po\UP{AA'i}e_{A'0})+ 3c_2\Vj\,e\UP{AA'i'}e_{A'0}\,
\sg\,0\,.
\end{split}
\end{equation}

\subsubsection{A relation between constraints }

In order to render the system of equations consistent,  one either
has to determine certain Langrange multipliers or to introduce new
constraints.  We will see that both steps will be necessary for
our system.

Performing the same calculation as the one leading to the second Bianchi 
identity\footnote{Spatial indices are raised according to convention (\ref{RaumindexHoch}).}
\[
 \DKovA R\UP{AB}|_{\HB_{123}}
 \,\id\,\DKovAK_i R\UP{ABi}\,=\,\partial_i R\UP{ABi}\,+
\, \Zshg\UP{A}_{Li} R\UP{LBi}\,-
\, \Zshg\UP{B}_{Li} R\UP{LAi}\,=0
\]
for the curvature with respect to the spatial components $\HB_{123}$
applied to the constraint $\XC\UP{ABi}$, eq.\ (\ref{DefXC}), yields
\begin{equation}
-\DKovAK_i  \XC\UP{ABi}\,=\,2\Va \DKovAK_i \Po\UP{ABi}+ \Vc \DKovAK_i 
e\UP{ABi}\,.
\end{equation}
Solving (cf eqs.\ (\ref{DefoCX}, \ref{DefoCY}, \ref{DefXC}, 
\ref{RaumindexHochBsp}))
\begin{equation}
-\oC\Up{AB}=-\oCX\UP{AB}-\oCY\Up{AB}  = 
 \DKovAK_i \Po\UP{ABi} +
 \Pe^{Ai} \xe\UP{B}_i- \Pe^{Bi} \xe\UP{A}_i \,,\\
\end{equation}
for $\DKovAK_i \Po\UP{ABi}$ and using the identity (\ref{EOM_omega}) 
for $S\Up{AB}=0$ one finds the six relations 
\begin{equation}\label{eq:ConsRel}
 \DKovAK_i  \XC\UP{ABi}\,=\,2\Va (\oC\UP{AB}+ 
\Pe\UP{Ai} e^B_i - \Pe\UP{Bi} e^A_i) +
\Vc(\YC\UP{Ai}\,e\UP{B}_i -\YC\UP{Bi}\,e\UP{A}_i)
\end{equation}
between the constraints $\XC\UP{ABi}$, $\oC\UP{AB}$ and $\YC\UP{Ai}$.

From (\ref{eq:ConsRel}) for invertible $\Va$ follow the six weak 
 relations (cf (\ref{DefoCY}))
\begin{equation}\label{eq:ePe}
J\UP{AB}:=-\oCY\UP{AB} = \Pe\UP{Ai} e^B_i - \Pe\UP{Bi} e^A_i \sg 0 \,
\end{equation}
between the $12$ momenta $\Pe\UP{Ai}$ and their conjugate variables $e_{Ai}$.
These relations exactly coincide with the constraints $J\UP{AB}$
in the previous literature \cite{DEIS76,HENE83}.  Although the
$\oC\Up{AB}$ appear naturally in our approach as the generators
of the local Lorentz transformations (and vanish weakly by
themselves), they may be tied as well by $\oC\UP{AB}\sg -
J\UP{AB}$ --- as can be observed from (\ref{eq:ConsRel}) --- to
these constraints which played that role in the usual tetrad
gravity.  However, in a theory where $\omega_{ABi}$ and $e_{Ai}$
are treated as independent variables as in our approach these are
not the generators of the local Lorentz transformations, but
only a part thereof.  This can be seen, for instance, by looking
at the Poisson brackets (\ref{PKoCXTC}) and (\ref{PKoCYTC}) of 
Appendix C 
where the \oQ non-Lorentzian\eQ\ parts 
just cancel in the sum of those two contributions to $\{ 
\Omega^{AB}, T^{A'i'} \}$. 

\subsubsection{Lagrange multipliers and ternary constraints}

The key to the solution of the constraint equation (\ref{XTimeEvolution}) is
 the observation that the terms in the parenthesis are just
 (cf (\ref{RaumindexHochBsp})) \footnote{A suggestive notation 
 for the projections $\partial_{0jk}\, \frac{\epsilon^{ijk}}{2} 
 = \vert^i$ and $\partial_{123}=\vert_{123}$ will be used from 
 now for the related components of 2-form and 3-form equations.
 \label{fnt14}}
\begin{equation}
	Y\UP{A}\wedge e\UP{B} - Y\UP{B}\wedge e\UP{A}|^i \sg 0 \,,
\end{equation}
and (\ref{eq:ePe}) is nothing else but 
\begin{equation}
	Y\UP{A}\wedge e\UP{B} - Y\UP{B}\wedge e\UP{A}|_{123} \sg 0 
\,.
\end{equation}
Thus one has to find a solution to 
\begin{equation}
	Y\UP{A}\wedge e\UP{B} - Y\UP{B}\wedge e\UP{A} \sg 0 
\end{equation}
which we already know to be uniquely $Y\UP{A}\sg0$ (cf lemma A of Appendix A):
\begin{eqnarray}
y_{Ai}& := & Y_{A0i}\,\sg\,0 \,, \label{DetYNull}\\
\Pe\Up{Ai}& := & \frac{1}{2} \epsilon\UP{ijk}Y_{Ajk}\,\sg\,0 \,. \label{ConPe}
\end{eqnarray}
Equation (\ref{DetYNull}) is a condition on Lagrange multipliers, 
whereas (\ref{ConPe}) are new (ternary) constraints. Actually, 
the result (\ref{DetYNull}) and (\ref{ConPe}) was to be 
expected by our analysis of the Lagrangian equations 
of motion (cf (\ref{YsNull})).

By means of these two conditions, however, not only 
$\dot{C}^{ABi}\approx 0$ (cf (\ref{XTimeEvolution})), 
but also $\dot{E}^A \approx 0$ 
(cf (\ref{eTimeEvolution})) is fulfilled automatically. 

Still, the conservation of the torsion constraint $\YC\UP{Ai}$
(\ref{YTimeEvolution}) and the one of the new momenta constraints
(\ref{ConPe}) (cf (\ref{PeTimeEvolution}) below)
 are not ensured yet.  This will provide 
conditions on the remaining Lagrange mulipliers $x_{ABi}$.

\subsubsection{Time evolution of $\Pe\UP{Ai}$}
The Poisson brackets of the ternary constraints (\ref{ConPe}) 
with all constraints are
\begin{align}
\PK{\Pe\UP{Ai}, \eC\UP{A'}} &\,=\,
  ( \Vc \,\Po\UP{AA'i}  + \widetilde{\Lambda}\Vj\,e\UP{AA'}\UP{i}) \dfs\,
 \label{PK:Pee} \\
\PK{\oC\UP{AB}, \Pe\UP{A'i'}} &\,=\, 
 (\eta\UP{AA'}\,\Pe\UP{Bi'} -\eta\UP{BA'}\,\Pe\UP{Ai'}) \dfs \,, 
\label{PK:oPe} \\
\PK{\Pe\UP{Ai}, \XC\UP{A'B'i'}} &\,=\,
\Vc\,(\eta\UP{AA'} e\UP{B'}_{k'}-\eta\UP{AB'} e\UP{A'}_{k'})
\,\eps\Up{ii'k'}\dfs \,,
\label{PK:PeX} \\
\PK{\Pe\UP{Ai}, \YC\UP{A'i'}} &\,=\, 
 -\eta\UP{AA'}\,\eps\Up{ii'k'}\dfs_{,k'} - 
 \Zshg\UP{AA'}_{k'}\,\eps\Up{ii'k'}\dfs \,,
  \label{PK:PeY} \\
\PK{\Pe\UP{Ai},\Pe\UP{Ai}} &\,= \,0 \,.
\end{align}
Taking into account eqs.\ (\ref{DetYNull}), (\ref{ConPe})
their time evolution is found to be
\begin{align}\label{PeTimeEvolution}
\dot{\Pe}\Up{Ai} &\, \sg\,
  e_{A'0} ( \Vc \,\Po\UP{AA'i}  + \widetilde{\Lambda}\Vj\,e\UP{AA'}\UP{i})\, +
\frac{1}{2}\,  x_{A'B'i'} \Vc\,(\eta\UP{AA'} e\UP{B'}_{k'}-\eta\UP{AB'} e\UP{A'}_{k'})
\,\eps\Up{ii'k'}  \notag\\
 &\, \sg\,
 \Vc ( e_{A'0}\Po\UP{AA'i} +  e_{B'k'}\, x\UP{AB'}\Do{i'} \,\eps\Up{ii'k'})
 + \widetilde{\Lambda}\Vj\, e_{A'0} e\UP{AA'}\UP{i}\,.
\end{align}
It should be noted that the right hand side of (\ref{PK:PeY}) is
the only instance where a derivative of the delta functional
occurs in our approach.  However, since the corresponding
Lagrange multipliers $y_{Ai}$ of the torsion constraints
$\YC\UP{Ai}$ vanish anyhow by (\ref{DetYNull}) and thus could be 
dropped in (\ref{PeTimeEvolution}), these particular Poisson
brackets are of no importance for the Hamiltonian analysis.

\subsection{The remaining Lagrange multipliers $x_{ABi}$}
\subsubsection{Reformulation of the problem}

Defining a 2-form
\begin{equation}
\label{eq:4.95}
Z^{AB} \equiv 2 \Va X^{AB} + \left(\Vc\, - 
\frac{\Lambda}{6}\right)\, e^{AB}\; ,
\end{equation}
we show in this subsection that the solution of $\dot{T}^{Ai} 
\approx 0$ (\ref{YTimeEvolution}) and $\dot{\pi}^{Ai} \approx 0$ 
(\ref{PeTimeEvolution}) can be reformulated as the solution of 
the two weak 3-form equations
\begin{eqnarray}
\label{eq:4.96}
Z^{AB} \wedge e_B &\approx& 0 \; ,\\
\label{eq:4.97}
\widehat{Z^{AB}} \wedge e_B & = & \Vj\, Z^{AB} \wedge e_B 
\approx 0 \; .
\end{eqnarray}
The components of $X^{AB}$ in (\ref{eq:4.95}) consist of the 
Lagrange multipliers $X_{AB0i} = x_{ABi}$ and of the 
${X^{AB}}_{jk}$ which are proportional to the momenta conjugate 
to ${\omega^{AB}}_i$ (cf (\ref{4.5})). Therefore, (\ref{eq:4.96}) 
and (\ref{eq:4.97}) will be inhomogeneous linear equations for 
$x_{ABi}$ whose solution will be discussed in subsection 4.3.2.

The derivation of each of the equations (\ref{eq:4.96}) and 
(\ref{eq:4.97}) consists of separate proofs for the projections 
$\vert^i$  and  $\vert_{123}$ (cf footnote \ref{fnt14}).
\noindent
We first note that the bracket in the second line of 
(\ref{YTimeEvolution}) is nothing else than $X^{AB} \wedge e_B  
\vert^i$.  Adding a vanishing term $(c_0 - \Lambda/6) \, e^{AB} 
\wedge e_B$ to the last term of (\ref{YTimeEvolution}), with $c_0 
+ \Vj c_2 = \Vc$ allows to identify that term with the 
component $\vert^i$ of the last two terms (\ref{eq:4.95}). This 
proves (\ref{eq:4.96})$\vert^i$. Also in 
(\ref{PeTimeEvolution}) the bracket in the second line is $X^{AB} 
\wedge e_B\, \vert^i$. Together with the prefactor $\Vc =  (\Vc 
\Va^{-1})\, \Va = - 4 (\gamma + \Vj)\, \Va$ (cf 
(\ref{3.12})) the combination $2\Va X^{AB} \wedge e_B\, \vert^i$ 
can be expressed by means of (the already proved) 
$(\ref{eq:4.96})\vert^i$ by the corresponding component from the 
last two terms in (\ref{eq:4.95}). Again we may add a vanishing 
term proportional $c_2\, e^{AB} \wedge e_B$ in order to produce 
the combination $c_0 + \gamma c_2 $ in (\ref{3.15}). Together 
with (\ref{3.16}) this leads (up to an overall factor ) to  
$(\ref{eq:4.97})\vert^i$. 

To show the validity of $(\ref{eq:4.96})\vert_{123}$ and  
$(\ref{eq:4.97})\vert_{123}$ the constraints (\ref{4.9}), 
(\ref{4.15}), (\ref{DefXC}) and (\ref{ConPe}) suffice. Indeed 
from (\ref{4.9}) with (\ref{ConPe}) one arrives at the same 
expression as the one in (\ref{PeTimeEvolution}) but projected 
upon $\vert_{123}$ instead of $\vert^i$. However, before 
concluding that $(\ref{eq:4.97})\vert_{123}$ holds, according to 
the intermediate step of our argument in the case $\vert^i$, we 
first need the proof of $(\ref{eq:4.96})\vert_{123}$. Multiplying 
the constraint (\ref{DefXC}) with $e_{Bi}$ one realizes 
immediately that the result in the expression would coincide with 
$(\ref{eq:4.96})\vert_{123}$ if $R^{ABi} \wedge e_{Bi} \approx 
0$. Taking the definition of torsion from the torsion constraint 
(\ref{4.15})  and  $R^{AB}$ from (\ref{DefR}) the second 
Bianchi identity in the form
\begin{equation}
\label{eq:4.98}
\partial_i T^{Ai} + \frac{\epsilon^{ijk}}{2} \, 
\omega^{AD}_j \, T_B^k = R^{ABi} e_{Bi}\; ,
\end{equation}
of course, also holds here which for (\ref{4.15}) implies the 
desired result. Thus (\ref{eq:4.96}) and (\ref{eq:4.97}) are 
valid for all components.

\subsubsection{Solution of $x_{ABi}$}
Each one of the two equations (\ref{eq:4.96}), (\ref{eq:4.97}) has 12 
components $\vert^i$ and four components $\vert_{123}$, yielding 
altogether 32 linear equations for the 18 Lagrangian multipliers 
$x_{ABi}$, 8 of which are identities not including $x_{ABi}$. 
Therefore, 6 relations between those components 
should hold to make a unique solution possible. 

As a first step one has to solve (\ref{eq:4.96}) and 
(\ref{eq:4.97}) as a system of 
linear inhomogeneous equations for $Z^{AB}_{0i}$, where the 
r.h.s.\ is linear in ${e^A}_0$. Then from the definition 
(\ref{eq:4.95}) 
the solution for ${x^{AB}}_i$ can be read off directly. This first 
step is by no means straightforward, however, it is enormously 
simplified by exploiting the invariance of (\ref{eq:4.96}) and 
(\ref{eq:4.97})  
which allows to perform that calculation in a suitable frame. 
Locally the components $e^A_I$ of the vierbeine can be 
transformed by  a Lorentz transformation ($\gamma = 
(1-v^2)^{-\frac{1}{2}}, v^2 = v_av^a$) 
\begin{equation}
\label{eq:99}
{\ell^A}_B \; = \; {\left(
\begin{array}{cc}
\gamma & \gamma\,v_{b}\\
\gamma v^a & \left[ \delta^a_b + (\gamma -1)\, \frac{v^av_b}{v^2}
\right]
\end{array}
\right)^A}_B
\end{equation}
and a diffeomorphism restricted to the space components $e_i^a$
\begin{equation}
\label{eq:100}
t_J{}^I = \left(
\begin{array}{cc}
1 & 0\\
0 & t_j{}^i
\end{array}
\right)\; ,
\end{equation}
such that
\begin{equation}
\label{eq:101}
\tilde{\tilde{e}}^A{}_J \; = \; {t_J}^I\, \tilde{e}^A_I = 
{t_J}^I \, \ell^A{}_B\, e^B_I
\end{equation}
is brought into the form
\begin{equation}
\label{eq:102}
\tilde{\tilde{e}}^{\ul{0}}_i = 0\; ,
\end{equation}
\begin{equation}
\label{eq:103}
\tilde{\tilde{e}}^a_i = \delta_i^a\; .
\end{equation}
Actually $t_j{}^i$ is nothing else than the 3d inverse 
${}^3\tilde{E}^i_a$ (in the sense ${}^3\tilde{E}^i_a\, \tilde{e}^b_i = 
\delta_a^b$) where the index $a$ is replaced by $j$. Physically 
(\ref{eq:102}) means that at the space time point considered, 
$\tilde{e}_i^0$, the analogon of the ``shift'', is set to zero 
and thus transformed to the rest frame, 
whereas the directions of the 3d holonomic frame have been 
rotated into the Lorentz directions renormalized to 
one.\footnote{This implies that in the present paper we restrict 
for simplicity to $e_I^A$ which are not light-like, a case which 
must be treated separately.} 
Eq.\ (\ref{eq:102}) with (\ref{eq:99}) implies  the velocity
\begin{equation}
\label{eq:104}
v_b = - {}^3E_b^i\, e_i^{\underline{0}}
\end{equation}
where now the 3d inverse of $e^a_j\;$ (in the sense ${}^3{E}^i_b\, e^b_j = 
\delta_j^i$) has been used. 

The condition $v^2 < 1$ for the local Lorentz transformation 
amounts to the requirement ${}^3g^{ij} e_i^{\ul{0}}\, 
e_j^{\ul{0}} < 1$ for the norm of $e_i^{\ul{0}}$ with respect to 
the local 3d metric. Light-like boosts are excluded in our 
present paper.

Clearly $\ell^A{}_B$, as well as its inverted form  
$\ell(-v)$ which appears, when one returns to the original 
frame, will depend in a complicated (nonpolynomial) way on the 
original $e^{\ul{0}}_i$ and $e_i^a$.

But the inverse $(t^{-1})_i{}^j$ simply coincides with 
$\tilde{e}^{(a)}_i$, where, as in the inverse ${}^3\tilde{E}^i_a$ 
above,  the index $(a)$ is replaced by the 
index $j$:\footnote{This mixed usage of some indices like $j$ now is 
indicated by $(j)$.}
\begin{eqnarray}
\label{eq:zab}
\tilde{Z}_{0i}^{ab} \; &=& \;  
\tilde{e}_i^{(j)}\; \tilde{\tilde{Z}}^{ab}_{0(j)}\; = \; 
{\ell^{(j)}}_B\, e^B_i\; \tilde{\tilde{Z}}^{ab}_{0(j)}\\
\label{eq:zac}
\tilde{Z}_{0i}^{ab} \; &=& \;  
\tilde{e}_i^{(j)}\; \tilde{\tilde{Z}}^{\ul{0}b}_{0(j)}\; = \; 
{\ell^{(j)}}_B\, e^B_i\; \tilde{\tilde{Z}}^{\ul{0}b}_{0(j)}\\
\label{eq:zad}
\end{eqnarray}
We recall that the Lorentz transformations in (\ref{eq:105}) and 
(\ref{eq:106}) depend on  (minus) the velocity (\ref{eq:104}). 
The components of $Z$ in the general frame (without tilde) 
\begin{equation}
\label{eq:zae}
Z^{AB}_{0j} \; = \; {(\ell^{-1})^A}_C\, (\ell^{-1})^B{}_D\, 
\tilde{Z}^{CD}_{0j}
\end{equation}
contain two more Lorentz transformations with (minus) the 
velocity (\ref{eq:104}). Therefore, at the end of the day, the 
factors in the 
relation between $Z$ and $\tilde{\tilde{Z}}$ are linear in 
$e^B_i$, except 
for the Lorentz transformations of the type (\ref{eq:99}) with 
(nonlinear) velocity (\ref{eq:104}) and, of course, a 
nonpolynomial dependence on $v$ in $\gamma$.

For foliations which do not require large velocities, a low 
velocity limit of (\ref{eq:99}) suffices. However, 
(\ref{eq:102}) cannot be reached in this way by a Galilei 
transformation. Instead for small velocities $v \ll 1$ also a transformation 
\begin{equation}
\label{eq:105}
\tilde{\ell}^A{}_B = {\left(
\begin{array}{cc}
1 & v^a \\
0 & \delta^a_b
\end{array}
\right)^A}_B\; ,
\end{equation}
may be considered which is a group contracted \cite{INOW53} 
Lorentz transformation, obtained by rescaling $v^a \to \lambda 
v^a, \, x^a \to \lambda^{-1} x^a$ in the limit $\lambda = 0$. 
This is a ``twisted'' version of the more familiar contraction towards the 
Galilei transformation ($v^a \to \lambda v^a, \, x^0 \to 
\lambda^{-1} x^0$). 

Having shifted essential parts of the problem into $\ell^a_b$ and $t$ 
the solution in the frame (\ref{eq:102}), (\ref{eq:103}) 
is relatively simple. Details are given in Appendix D (in the 
frame (\ref{eq:102}), (\ref{eq:103} ) holonomous and 
anholonomous spatial indices may be raised, lowered and combined 
freely):
\begin{equation}
\label{eq:106}
\tilde{\tilde{Z}}^{\ul{0},a}_{0i} \; =\;
-\frac{1}{4} \epsilon^{aB}{}_{CD} \, 
\tilde{\tilde{Z}}^{CD}_{jk} \,
\tilde{\tilde{e}}_{B0}\, \epsilon^{ijk}
\end{equation}
\begin{equation}
\label{eq:107}
\tilde{\tilde{Z}}^{ab}_{0i} \; =\;
- \frac{1}{2}\, \epsilon^{abj}\, 
\tilde{\tilde{Z}}^{\ul{i}B}_{rs}\, \epsilon^{jrs}\, 
\tilde{\tilde{e}}_{B0} 
\end{equation}
Together with the vanishing multipliers (\ref{DetYNull}) this resolves the 
problem.  As expected, the solution is found to be the 
consequence of 18 among the 32 equations (\ref{eq:4.96}), 
(\ref{eq:4.97}), the remaining ones just yielding identities. The 
expressions on the \mbox{r.h.s.-s} of (\ref{eq:106}) and (\ref{eq:107}) 
not only are linear in $\tilde{\tilde{e}}_0^a$  but also in 
the canonical momenta $\tilde{\tilde{\Pi}}^{ABi}$ (cf 
(\ref{RaumindexHochBsp})) 
proportional to $ \tilde{\tilde{Z}}^{ABi}_{jk}$. These 
properties survive the (only $e_i^A$--dependent) linear 
transformations $t^{-1}, \, \ell (-v)$ when returning 
by (\ref{eq:105}) -- (\ref{eq:107}) to the 
original frame $e_I^A$.

Thus the Lagrange multipliers are expressible as
\begin{equation}
\label{eq:ab}
 x_{ABi} = {h_{ABi}}^C\, e_{C0}
\end{equation}
where the coefficients ${h_{ABi}}^C = - {h_{BAi}}^C $ 
are nonpolynomial functions of $e_i^A$, but still linear in $\Pi^{ABi}$. 
According to (\ref{Hcan}) with (\ref{DetYNull}) and (\ref{eq:ab}) 
this yields a 
Hamiltonian of the form (cf also the last ref.\ \cite{HENE83}) 
\begin{equation}\label{HamAnsatz}
        H = \int \, d\Up{3}x \,\left(
	 e_{A0} \tilde{E}\UP{A} + \Bruch{1}{2} \xo_{AB0} \oC\UP{AB} \right)\,,
\end{equation}
with the new constraints 
\begin{equation}
\label{newEAnsatz}
 \tilde{E}\UP{A} \,=\, E\UP{A} + \frac{1}{2} \, {h_{CDi}}^{A}\, 
 C^{CDi}\; , 
\end{equation}
which must be first class by comparing with the number of
degrees of freedom: Our theory involves the $30$ phase space
variables $(e_{Ai}, \omega_{ABi})$, the $N_f=10$ first class
constraints $(\tilde{E}\Up{A}, \oC\Up{AB})$ or equivalently the
$10$ independent Lagrange multipliers $(e_{A0}, \omega_{AB0})$,
the $12+18+12=42$ second class constraints $(\YC\Up{Ai},
\XC\Up{ABi}, \pi\Up{Ai})$ with the six relations
(\ref{eq:ConsRel}) between them,  
giving $N_s=36$ independent second class
constraints.  Thus, we find the number of physical degrees of
freedom according to standard textbook lore (cf e.g.\
\cite{HT92}) as
\begin{equation}
 N_{df} \,= \textrm{N}_{ph} - N_f - \frac{1}{2} N_{s}\,=
            \, 30 - 10 - \frac{1}{2}(42-6) = 2\,,
\end{equation}
which is precisely the correct one for GR and thus agrees 
with counting according to  our Hamiltonian analysis.

It should be noted that independent of the specific variables one 
always encounters 10 first class constraints: In Einstein-Cartan 
gravity with 12 variables $e_{Ai}$ one obtains $12 - 10 = 2$ 
degrees of freedom, the Ashtekar approach with 18 $\omega_{ABi}$ 
the counting is $18-10-\frac{12}{2}=2$,  where the last term 
originates from the second class constraints in that formulation.

The algebra of first class constraints (with respect to 
appropriate Dirac brackets) should
(strongly) equal the Poincar\'{e} algebra:\footnote{Of course, 
relations of this form, with the exception of (\ref{PoinC3}),
have been conjectured from symmetry arguments for a long time.  
They are e.g.\ the 
basis of Poincar\'e gauge theory \cite{KIBB61,UTI56,HKN76}     
(cf also the last reference \cite{HENE83}).
Primary constraints with such an algebra also appeared in 
refs.\ \cite{CHAR83} in a tetrad theory with dependent spin 
connection. However, in contrast to the results of that works, as 
seen from (\ref{PK:oPe})  our Lorentz constraints possess 
\textit{all} the correct commutators, especially also the ones 
with the momenta $\pi^A_i$ of the vierbeine.}
\begin{align}
\PK{\oC\UP{AB}, \oC\Up{A'B'}} &\,=\, (\eta\UP{AA'}\oC\UP{BB'} -
\eta\UP{BA'}\oC\Up{AB'}+\eta\UP{BB'}\oC\Up{AA'}-
\eta\UP{AB'}\oC\Up{BA'})\delta\,,
\label{PoinC1} \\
\PK{\oC\UP{AB}, \efC\UP{A'}}& \,=\, (
\eta\UP{AA'}\efC\Up{B}-\eta\UP{A'B}\efC\Up{A})\delta\,, \label{PoinC2} \\
\PK{\efC\UP{A}, \efC\Up{A'}} &\,=\, 0 \; . \label{PoinC3}
\end{align}
So far this final check 
has not been made, because of the very messy algebra.

\section{Conclusion and Outlook }

Motivated by the success of an analogous program in 
$1+1$ dimensional gravity theories we present a reformulation of 
Einstein (-deSitter) gravity, strictly in terms of Cartan 
variables (one-form vierbeine $e_A$ and spin connection 
$\omega_{AB}$), and of a first order derivative Hamiltonian  
action. We  do not rely upon the Palatini mechanism to arrive at the 
dependent spin connection, but keep a separate condition for 
vanishing torsion. For that and in order to reproduce usual 
Einstein gravity, sets of auxiliary two-form  fields 
($X^{AB}, Y^A$) 
are introduced. Part of them become the momenta, canonically 
conjugate to the dynamical components of (space-like parts) of 
spin connection and of the vierbeine. The remaining components of 
the auxiliary fields represent Lagrangian multipliers. 

By comparison with the situation in $1+1$ dimensional gravity, 
resulting from spherical reduction of Einstein gravity in four 
dimensions, the dilaton field $X$ in that case may even have a 
counterpart among  certain ``dilaton fields'' contained in 
$X^{AB}$ which also here may be the result of a compactification 
mechanism from gravity in dimensions larger than four. 

Another speculation suggests itself by the different ways the two 
contributions to the effective cosmological constant $\Lambda$ in 
(\ref{3.16}) appear in the present formulation. By now 
it has been 
established by a host of astronomical data \cite{PERL99} that 
$\Lambda$, the dark energy, has a small but positive value. 
This disagrees with supergravity which requires an Anti-deSitter 
space with negative $\Lambda$ \cite{TOWN77}. That this result is 
unchanged in related dilaton supergravity, e.g.\ for the $1+1$ 
dimensional case has been demonstrated in \cite{BGK04}. 
Assuming that the parts (\ref{ActLDSX}) and (\ref{ActLDSY}) 
of the action by themselves are the result of compactifying a 
higher dimensional supergravity, whereas the contribution 
(\ref{ActLDSCC}) is determined by the positive contribution from 
the Standard Model, there is a chance of compensation, albeit 
still suffering from the usual tremendous fine-tuning problem.  
In any case, the undetermined sign of the second term in (\ref{3.16}) 
allows such a solution.

In our approach we do not 
follow the usual route of an ADM decomposition, instead being led 
rather naturally  to a Hamiltonian analysis, reminiscent of the 
one in nonabelian gauge theory. Nevertheless, our approach 
completely differs from the one advocated by Ashtekar 
\cite{ASHT86}  
and the work which has developed from that. On the other hand, in 
the course of our analysis we encounter constraints which in 
classical works on the interpretation of the e.g.\ Lorentz 
invariance had been introduced --- to a varying degree --- as ad 
hoc conditions (cf e.g.\ refs.\ \cite{DEIS76,CHHE88}).  

The price we have 
to pay is the appearance of ternary and second class constraints 
which determines the aforementioned Lagrange multipliers, 
the ternary constraints being nothing else than the momenta 
canonically  
conjugate  to the space components of the vierbeine. 
Already in order to arrive at this point required the evaluation 
of straightforward but --- at intermediate steps --- very lengthy 
formulas for Poisson brackets which has been done by hand and 
verified on the computer by a program package written for this 
purpose. 

The very symmetric structure of our approach allowed us to 
formulate the final Hamiltonian in
a manner akin to nonabelian gauge theories, i.e.\ in terms of
first class constraints (with respect to appropriate Dirac 
brackets), multiplied by ``time'' components
($e_{A0}$) of the vierbeine and of the spin connection
($\omega_{AB0}$) --- just like the Gauss constraints in
Yang-Mills theories.  The explicit check of the first class 
property for the set $\Omega^{AB},\, \Tilde{E}^A$ still requires 
another exacting mathematical effort, but we are convinced that the 
constraints 
of the (local) Poincar\'{e} algebra (\ref{PoinC1})--(\ref{PoinC3}) 
should be reproduced directly.

At this point we should emphasize again that our approach 
contains the original SO(3,1) local Lorentz symmetry with 
associated connection --- not reduced to SO(3,$\mathbb{C}$) as in 
Ashtekar's formulation.  In comparison with other manifest 
SO(3,1) approaches (cf e.g.\ \cite{Alexandrov,Alex+Vas}) our 
constraints and the associated constraint algebra are much less 
involved.

Our approach lends itself to generalisations\footnote{For a 
comprehensive review of generalized gravity cf \cite{HMMN95}} in 
several directions, the most obvious one being to dynamical 
torsion which also plays a role in the attempts to make GR 
renormalizable (cf e.g.\ \cite{SEZ80}). Then an ansatz like 
(23)--(25) must be supplemented by further terms linear and 
quadratic $Y^A$. Dynamical torsion is a natural consequence of 
Poincar\'e gauge theories of gravity 
\cite{KIBB61,UTI56,HKN76}.~\footnote{\mbox{
A review on torsion, covering in particular the 
literature on actions quadratic in torsion is \cite{HAM02}.}}

Another interesting field of applications may be the 
teleparallelism formulations of Einstein gravity where the spin 
connection is flat and the whole dynamical structure resides in 
the Weitzenb\"{o}ck connection which is determined by torsion 
alone \cite{CART79} (cf also  here the review \cite{HAM02}).

However, it should be kept in mind that a polynomial ansatz with 
auxiliary fields as in our case places certain restrictions upon 
quadratic terms in torsion and --- for that matter --- also upon 
terms quadratic in the curvature tensor (cf the remarks after 
eqs.\ (\ref{DefGaussBonnet}), (\ref{DefPontrijagin})). 
On the other hand, the extension of the space 
of variables by auxiliary fields may be helpful in a quantum 
theory. Of course, it cannot  be expected that a full background 
independent quantization of gravity can be obtained as in $1+1$ 
dimensions \cite{KULI97,REPO02}. But our approach, although different 
from Ashtekar's one \cite{ASHT86}, may allow too some of the 
developments which were made possible by its structural 
similarity to  Yang-Mills theory in the Hamiltonian analysis.

\section*{Acknowledgement:}
The authors are grateful to S.\ Deser and P.\ van Nieuwenhuizen 
for correspondence and thank D.V.\ Vassilevich for helpful 
comments. 
In previous stages of this work one of us (H.S.) has been 
supported by Jubil\"{a}umsfondprojekt Nr.\ 7304 of the Austrian National Bank.

\newpage

\appendix

\renewcommand{\theequation}{\thesection.\arabic{equation}}
\numberwithin{equation}{section}

\section{Form equations}
In the analysis of the four-dimensional  Hamiltonian linear derivative action 
one frequently encounters three-form equations of the type 
\begin{equation}\label{XwedgeE}
Z\UP{A}\wedge e\UP{B} - Z\UP{B}\wedge e\UP{A} = S\UP{AB}\, \quad A, B \,=\, 0..3\,
\end{equation}
for the two-form components $(Z\UP{A})$ of a $(1,2)$ bi-form\footnote{
The notion of a bi-form proves to be very useful in
mathematics on manifolds.  In short, a bi-form is a tensor
product of alternating forms and alternating vectors, \ie\
alternating vectors whose components are forms.  Thus, for
example, a $(1,2)$ bi-form $Z=Z_{A} E\UP{A}$ is a vector which
two-form components $Z_A$ with respect to the basis $E\UP{A}$. 
Further useful abbreviations are $(e^2)^{AB} = e^{AB} = e^A \wedge e^B,\,  
(e^3)^{ABC} = e^{ABC} = e^A\wedge e^B \wedge e^C$ etc.
} for the components $(S\UP{AB})$ of an arbitrary right hand
side (2,3)-form.\\
{\bf Lemma A:} 
Equations (\ref{XwedgeE}) have a unique solution 
(cf Appendix A of  \cite{HSDr02}) 
\begin{equation}
Z\UP{A} \,= \, \Dinn_K S\Up{AK}-
\,\Bruch{1}{4}\, \Dinn_L \Dinn_K S\Up{KL}\Keil e\Up{A}\, , 
\end{equation}
with the inner derivative $\Dinn_K$ meaning evaluation of the
form on the basis vector $E_K$.

In our present paper we mostly need the homogeneous case $S^{AB} 
= 0$. Here clearly one solution is the trivial one 
$Z\UP{A}=0$. A direct proof for the uniqueness of this solution to 
(\ref{XwedgeE}) with $S^{AB}=0$ follows from  
considering the equations for the components $Z\UP{A}\Do{CD}$ in 
$Z^A = Z^A_{CD}\, e^{CD}$ which follow from (\ref{XwedgeE}) 
at $S^{AB} = 0$
\begin{equation}
 Z\UP{A}_{CD} e\UP{CDB} = Z\UP{B}_{EF} e\UP{EFA}\,\quad A\neq B\;.
\end{equation}
This gives for $A=0, B=1$  the three-form equation\footnote{
Here for notational simplicity we deviate from our convention to 
underline special Lorentz components.}
\begin{equation}
 Z\UP{0}\Do{23} e\UP{231} +
 Z\UP{0}\Do{02} e\UP{021} +
 Z\UP{0}\Do{03} e\UP{031} 
 \,=\, 
 Z\UP{1}\Do{23} e\UP{230} +
 Z\UP{1}\Do{12} e\UP{120} +
 Z\UP{1}\Do{13} e\UP{130}\, ,
 \end{equation}
from which one immediately infers 
$Z\UP{0}_{23}=0,Z\UP{1}_{23}=0$, as well as $(e\UP{021}=-e\UP{120})$
\begin{align}
 Z\UP{0}\Do{02} & \,=\, - Z\UP{1}_{12}\,, \\
 Z\UP{0}\Do{03} & \,=\, - Z\UP{1}_{13} \,. 
\end{align}
One sees that they are of the form $Z\UP{A}_{AB}=-Z\UP{C}_{CB}$
for $A\neq C$.  Thus considering all three equations ($A\neq B$
with $A,B=0..3$) for the components $Z\UP{A}\Do{AB}$ one finally
has (without loss of generality we may set e.g.\ $B=2$)
\begin{equation}
 Z\UP{0}\Do{02} \,=\, - Z\UP{1}_{12} \,=\,  Z\UP{3}_{32} 
 \,=\, - Z\UP{0}\Do{02} \Implies Z\UP{0}\Do{02}\,=\,0 \; ,
\end{equation}
and thus the trivial solution $Z^A = 0$ is the unique one.

\section{Hat calculus}

If we consider the Hat operator as acting not only on 
bi-forms but  on any antisymmetric object $X_{AB}$
with two $SO(3,1)$ indices, using the definition 
\begin{equation}
 \myHat{X_{AB}}  \,:=\, \Halb \eps_{AB}{}^{CD} X_{CD}\, ,
\end{equation}
 and raising and lowering indices with the Minkowski metric $\eta_{AB}$ 
we find the following rules:\\
%


{\bf Lemma B (Hat rules):} 
Let $\diamond$ denote any bilinear operator and $X_{AB}, Y_{AB}$
be any antisymmetric object with indices $A,B=0,1,2,3$ then
following identities hold:
\begin{align}
\myHat{\myHat{X_{AB}}}&\,=\,-X_{AB}\,,\label{eq:Hat:1}\\
\myHat{X_{KL}}\diamond Y^{KL}&\,=\,
X_{KL}\diamond\myHat{Y^{KL}}\,,\label{eq:Hat:2}\\
\myHat{X_{AL}\diamond Y^{L}{}_B} &\,=\, \Halb \left( 
\myHat{X_{AL} \diamond Y^{L}{}_B}-\myHat{X_{BL} \diamond Y^{L}{}_A}\right)\,,\label{eq:Hat:3}\\
\myHat{X_{AL}\diamond Y^{L}{}_B} &\,=\, \Halb \left( 
\myHat{X_{AL}} \diamond Y^{L}{}_B-\myHat{X_{BL}} \diamond Y^{L}{}_A\right)\,,\label{eq:Hat:4}\\
\myHat{X_{AL}} \diamond Y^{L}{}_B &\,=\,-X_{BL} \diamond \myHat{Y^{L}{}_A}\,,
\quad A\neq B\,,\label{eq:Hat:5}\\
\myHat{X_{AL}} \diamond \myHat{Y^{L}{}_B} &\,=\,
\Halb\eta_{AB}X_{KL} \diamond Y^{KL} \,+\,X_{BL} \diamond Y^{L}{}_A\,.\label{eq:Hat:6}
\end{align}

Of course, this Hat operation on operands with antisymmetric
indices is again linear.  Moreover it commutes with the exterior
derivative because $\eps_{ABCD}$ and $\eta_{AB}$ are merely
numbers, yielding the identities
\begin{align}
 \myHat{\mu X_{AB}+\nu Y_{AB}}&\,=\, 
 \mu\myHat{X_{AB}}+\nu\myHat{Y_{AB}}\; , 
 \label{eq:Hat:lin} \\
 d \myHat{Z_{AB}} &\,=\, \myHat{d Z_{AB}}\; .
 \label{eq:Hat:d} 
\end{align}
Covariant derivatives on bi-vectors are defined by
\begin{equation}
\label{eq:Hat:e}
(DX)^{AB} = dX^{AB} + {\omega^A}_C \wedge X^{CB} + {\omega^B}_C 
\wedge X^{AC}
\end{equation}

\section{Poisson brackets}

Our Poisson brackets are defined  as
\begin{equation}
\{ A\, (x^0,x^i), \, B(x^0, {x^\prime}^i)\, \} = 
\int\, d^3 \tilde{x}\, \left\{ \frac{\delta A}{\delta\, 
\tilde{e}_{Ai}}\; \frac{\delta B}{\delta \tilde{\Pi}^{Ai}} + 
\frac{1}{2}\, \frac{\delta A}{\delta \tilde{\omega}_{ABi}}\, 
\frac{\delta B}{\delta \tilde{\Pi}^{ABi}} - (A\leftrightarrow 
B)\; \right\}\, .
\end{equation}
An expression like $\tilde{e}_{Ai}$ is the shorthand for $e_{Ai} 
(x^0,\tilde{x}^i)$ etc. Note the factor $\frac{1}{2}$ in the 
second term which permits independent summation of $A$ and $B$ in 
the antisymmetric components. 

Another simple formula which is crucial in order to obtain 
brackets of constraints which do not produce derivatives of 
$\delta$-functions, and in this sense resembling the ones in Yang-Mills 
theory, is 
\begin{equation}
\int_{x^\prime} \, \left[\, f(x')\, \delta (x^{\prime\prime}-x) \, \partial^\prime\, 
\delta (x^\prime-x^{\prime\prime}) + (x' \leftrightarrow x ) \, \right] = 
(\partial f)\, \delta (x^\prime-x)\; .
\end{equation}
In order to give an idea of the brackets to be evaluated, a few 
examples are given below for different portions of $\Omega^{AB}$ 
and $E^A$, as defined in (\ref{DefoCX}) -- (\ref{4.14}). In all 
cases the terms in 
the bracket again are taken at $(x^0, x^i)$ and $(x^0, {x^\prime}^i)$, 
respectively; $\delta\, (x^i-{x^\prime}^i) = \delta,\; \delta_{,i} = 
\partial_i\,\delta$: 
\begin{equation}
\begin{split}
\left\{\Omega^{AB}_{(X)}, \aVec \Po\UP{A'B'i'}\right\}\,=\, 
\aVec( &\eta\UP{AA'} \Po\UP{BB'i'} -\eta\UP{BA'} \Po\UP{AB'i'}
+\\[1mm]&
\eta\UP{BB'} \Po\UP{AA'i'}-\eta\UP{AB'} \Po\UP{B'A'i'})\, \delta
\end{split}
\end{equation}
\begin{equation}\label{ap:PB:oX1}
\begin{split}
\left\{\Omega_{(X)}\UP{AB}, R\UP{A'B'i'}\right\}\,=\, 
-( &\eta\UP{AA'} R\UP{BB'i'} -\eta\UP{BA'} R\UP{AB'i'}
+\\[1mm]&
\eta\UP{BB'} R\UP{AA'i'}-\eta\UP{AB'} R\UP{BA'i'})\, \delta
\end{split}
\end{equation}
\begin{equation}\label{ap:PB:oe11}
\begin{split}
\left\{\,\Omega_{(X)}\UP{AB}, E_{(X)}\UP{A'}\right\}\,=\, &
(\eta\UP{AA'} E^B -\eta\UP{BA'} E^B) \,\delta +\,\\[1mm] &
\cVec(e\UP{A}_i\, \Po\UP{BA'i}  - e\UP{B}_i\, \Po\UP{AA'i})\, 
\delta 
\end{split}
\end{equation}
\begin{equation}
\begin{split}
\left\{\Omega_{(X)}\UP{AB}, E_{(Y)}\UP{A'}\right\}\,=\, &
(\eta\UP{BA'} \Pe\UP{Ai}-\eta\UP{AA'} \Pe\UP{Bi})\,\delta_{,i} +\\[1mm]
&(\eta\UP{BA'}\, E^A_{(Y)}-\eta\UP{AA'}\,  E^B_{(Y)})\,\delta+
\\[1mm]&
( \omega\UP{BA'}_i \Pe\UP{Ai} -  \omega\UP{AA'}_i \Pe\UP{Bi})\, 
\delta  
\end{split}
\end{equation}
\begin{equation}
\begin{split}
\left\{ \Omega_{(Y)}\UP{AB}, E_{(X)}\UP{A'}\right\}\,=\,&
\cVec (\Po\UP{AA'i} e\UP{B}_i -\cVec \Po\UP{BA'i} e\UP{A}_i) \, 
\delta
\end{split}
\end{equation}
\begin{equation}
\begin{split}
\left\{\Omega_{(Y)}\UP{AB}, E_{(X)}\UP{A'}\right\}\,=\, &
(\Zshg\UP{AA'}_i \Pe\UP{Bi}-\Zshg\UP{BA'}_i \Pe\UP{Ai})\, \delta \\[1mm]&
(\eta\UP{AA'}\Pe\UP{Bi}-\eta\UP{BA'}\Pe\UP{Ai}) \delta_{,i}\, 
\delta
\end{split}
\end{equation}
\begin{equation}\label{PKoCXTC}
\begin{split}
\PK{\oCX\UP{AB}, \YC\UP{A'i'}}\,=\, &
(\eta\UP{AA'}\YC\UP{Bi'}-\eta\UP{BA'}\YC\UP{Ai'})\, \delta+
\\[1mm] &
(\eta\UP{BA'}e\UP{A}\Do{k'}-\eta\UP{AA'}e\UP{B}\Do{k'}) \epsIs  
\delta_{,j'} +
\\[1mm] &
(\Zshg\UP{BA'}_{j'}\, e\UP{A}_{k'} - \Zshg\UP{AA'}_{j'}\, e\UP{B}_{k'}) 
\,\epsIs\, \delta
\end{split}
\end{equation}
\begin{equation}\label{PKoCYTC}
\begin{split}
\PK{\oCY\UP{AB}, \YC\UP{A'i'}}\,=\, &
(\eta\UP{AA'}e\UP{B}\Do{k'}-\eta\UP{BA'}e\UP{A}\Do{k'}) \epsIs  
\delta_{,j'} +
\\[1mm] &
(\Zshg\UP{AA'}_{j'}\, e\UP{B}_{k'} - \Zshg\UP{BA'}_{j'}\, e\UP{A}_{k'}) 
\epsIs\, \delta
\end{split}
\end{equation}

\section{Solution for Lagrangian multipliers}

The solution of (\ref{eq:4.96}) and (\ref{eq:4.97}) is most easily 
established in the frame (\ref{eq:102}), (\ref{eq:103}), 
dropping double tildes in this Appendix:
\begin{equation}
\label{eq:daa}
e^{\ul{0}}_i = 0, \qquad e^a_i = \delta_i^a
\end{equation}
Projecting (96) and (97) upon $\partial_{0ij}$, respectively, 
yields 24 inhomogeneous equations for the 18 variables 
$Z^{AB}_{0j}$
\begin{eqnarray}
\label{eq:dba}
Z^{AB}_{0j}\, e_{Bk}\, \epsilon^{ijk} \; & = & 
-\frac{1}{2} \, Z^{AB}_{jk} \, \epsilon^{ijk} \, 
e_{B0} \equiv r^{Ai}\, ,\\
\label{eq:dbb}
\hat{Z}^{AB}_{0j}\, e_{Bk}\, \epsilon^{ijk} \; & = & 
-\frac{1}{2} \, \hat{Z}^{AB}_{jk} \, \epsilon^{ijk} \, 
e_{B0} \equiv \hat{r}^{Ai}\, ,
\end{eqnarray}
where the r.h.s.-s are linear in the momenta $\Pi^{ABi}$ (cf 
(65), (66), (95)) and in $e^A_0$.

The projections by $\partial_{123}$ imply 8 restrictions upon the 
18 momenta canonically conjugate to $\omega_{iAB}$
\begin{eqnarray}
\label{eq:dca}
Z^{AB}_{ij} \, e_{Bk}\, \epsilon^{ijk} \; & = &\; 0\; ,\\
\label{eq:dcb}
\hat{Z}^{AB}_{ij} \, e_{Bk}\, \epsilon^{ijk} \; & = &\; 0\; ,
\end{eqnarray}
which is reasonable because for $\pi_e^{Ai} \approx 0$ the 
effective phase space should be of dimension 20.

Multiplying (\ref{eq:dca}) and (\ref{eq:dcb}) by $e_{A0}$ 
verifies two relations between the $r^{Ai}$ and $\hat{r}^{Ai}$:
\begin{eqnarray}
\label{eq:dda}
r^{Ai}\, e_{Ai} \; & = & \; 0 \\
\label{eq:ddb}
\hat{r}^{Ai}\, e_{Ai} \; & = & \; 0 
\end{eqnarray}
In the frame  (\ref{eq:daa}) eqs.\ (\ref{eq:dba}) and 
(\ref{eq:dbb}) are each separated into the ones for $A=\ul{0}$ 
and $A=a$ (position and combinations of special indices here may 
be moved freely)
\begin{eqnarray}
\label{eq:dea}
Z^{\ul{0}a}_{0j}\, \epsilon^{ija} \; & = & \; r^{\ul{0}i}\, ,\\
\label{eq:deb}
\hat{Z}^{\ul{0}a}_{0j}\, \epsilon^{ija} \; & = & \; \hat{r}^{\ul{0}i}\, ,
\end{eqnarray}
and the eighteen equations for the eighteen $Z^{AB}_{0j}$
\begin{eqnarray}
\label{eq:dfa}
\left( \, M^{ij}\right)^a \, Z^{\ul{0}a}_{0j} \; &=& \; 
\hat{r}^{ai}\, ,\\
\label{eq:dfb}
-\left(\, M^{ij}\right)^a \, \hat{Z}^{\ul{0}a}_{0j} \; &=& \; 
r^{ai}\, ,
\end{eqnarray}
where for symmetry reasons $\hat{Z}^{\ul{0}a} = \frac{1}{2} 
{\epsilon^{\ul{0}a}}_{bc}\, Z^{bc}$ have been introduced as a 
second triplet of $Z$-s. Inspection of the matrices
\begin{equation}
\begin{array}{cccccc}
\label{eq:dga}
(M^{ij})^{\ul{1}}&=\quad (&\; 0 \quad &\quad \epsilon^{3ij}  \quad
&\quad - \epsilon^{2ij}\;&) \\
(M^{ij})^{\ul{2}}&=\quad (&\;- \epsilon^{3ij}  \quad
&\quad 0 \quad & \epsilon^{1ij}\;&)\\
(M^{ij})^{\ul{3}}&=\quad (&\;\epsilon^{2ij}\quad  &\quad  - 
\epsilon^{1ij}\quad &\quad 0\;&) 
\end{array}
\end{equation}
in (\ref{eq:dfa} and (\ref{eq:dfb}) shows that $Z^{ab}_{0j}$ for 
$b\neq i$ only have one  entry in the matrix $M$, so that   e.g.\ 
for (\ref{eq:dfa}) as long as $b\neq i$ 
\begin{equation}
\label{eq:dha}
Z^{ab}_{0i} \; = \; - \epsilon^{abj}\, \hat{r}^{\ul{i}j}\; .
\end{equation}
On the other hand, for $(b) = (i)$ (this notation indicates no 
sum over those indices) e.g.\ from (\ref{eq:dba}) the set of 
linear equations
\begin{equation}
\label{eq:dia}
Z^{\ul{0}(a)}_{0(a)} \; = \; \frac{1}{2}\, {A^a}_b \, 
\hat{r}^{\ul{b}b}\; ,
\end{equation}
\begin{equation}
\label{eq:dja}
{A^a}_b \quad = \quad 
{\left(\; 
\begin{array}{ccc}
1 & -1 & -1 \\
-1 & 1 & -1 \\
-1 & -1 & 1
\end{array}\right)^a}_b\; .
\end{equation}
follows. However, thanks to (\ref{eq:ddb}) for unit matrix 
$e_{ai}$ (\ref{eq:daa}) we may use $\hat{r}^{\ul{a}a}=0$ so that 
(\ref{eq:dha}) holds as well for equal indices $b=i$. Therefore, 
with double tilde restored (\ref{eq:dha}) becomes (\ref{eq:107})
of Section 4.3.2, and (\ref{eq:dfa}) by the same steps with 
\begin{equation}
\label{eq:dka}
Z^{\ul{0}a}_{0i} \; = \; \hat{r}^{ai}
\end{equation}
leads to (\ref{eq:106}).

So far the searched for 18 variables have been determined from 18 
equations. As expected from the rank of the system the still 
unused eqs.\ (\ref{eq:dea}) and (\ref{eq:dfa}) indeed turn out to 
be satisfied identically. 

In view of the complicated dependence of the velocity in 
(\ref{eq:99}) one could avoid that transformation in the 
transition from $Z$ to $\tilde{\tilde{Z}}$ by dropping the first 
condition (\ref{eq:daa}). The strategy of the solutions to 
(\ref{eq:dba}) and (\ref{eq:dbb}) in principle remains the same 
as indicated above, but, of course, the solution of the linear 
system is more involved.


\end{document}

%% file: smac.tex
\def\ul{\underline}
\newcommand{\oquote}{``}%
\newcommand{\equote}{''}%
\newcommand{\Up}[1]{^{#1}}%
\newcommand{\UP}[1]{^{#1}{}}
\newcommand{\DO}[1]{_{#1}{}}%
\newcommand{\Do}[1]{_{#1}}
\newcommand{\eps}{\epsilon}
\newcommand{\epsIs}{\,\eps\UP{i'j'k'}\,}
\newcommand{\HB}{\partial}
\newcommand{\myVec}[1]{\mathfrak{#1}}
\newcommand{\aVec}{\myVec{a}}
\newcommand{\cVec}{\myVec{c}}

\newcommand{\ie}{i.e.}
\newcommand{\EOMs}{e.o.m.-s}%
\newcommand{\Complex}{\mathbb{C}}

\newcommand{\Implies}{\Rightarrow}
\newcommand{\id}{\equiv}
\newcommand{\sg}{\approx}

\newcommand{\Keil}{\wedge}
\newcommand{\Bruch}[2]{ \frac{\displaystyle #1}{\displaystyle #2} }
\newcommand{\myHat}[1]{\widehat{#1}}
\newcommand{\Hodge}{\ast}
\newcommand{\DKovA}{\mathrm{D}}
\newcommand{\DKovAK}{\mathcal{D}}
\newcommand{\Dinn}{\iota}
\newcommand{\PK}[1]{\{#1\}}
\newcommand{\CA}[1]{\mathbf{#1}}
\newcommand{\CAi}[1]{#1\!\Do{\mathbf{(1)}}}
\newcommand{\CAii}[1]{#1\!\Do{\mathbf{(2)}}}

\newcommand{\Halb}{\ensuremath{\Bruch{1}{2}}}%
\newcommand{\HalbF}[1]{\Bruch{#1}{2}}%

\newcommand{\Tor}{\tau}
\newcommand{\Zshg}{\omega}

\newcommand{\PHNum}[1]{\mathbf{#1}}

\newcommand{\eC}{E} 
\newcommand{\efC}{\tilde{E}} 
\newcommand{\oC}{\Omega} 
\newcommand{\oCX}{\Omega\Do{(X)}} 
\newcommand{\oCY}{\Omega\Do{(Y)}} 
\newcommand{\eCX}{E\Do{(X)}} 
\newcommand{\eCY}{E\Do{(Y)}} 
\newcommand{\XC}{C{}} 
\newcommand{\YC}{T{}} 

\newcommand{\Pe}{\pi} 
\newcommand{\Po}{\Pi} 
\newcommand{\xe}{e} 
\newcommand{\xo}{\omega} 
\newcommand{\Ko}{x} 
\newcommand{\dfs}{\delta}
\newcommand{\epsI}{\,\eps\UP{ijk}\,}

\newcommand{\Va}{\mathbf{a}}
\newcommand{\Vb}{\mathbf{b}}
\newcommand{\Vc}{\mathbf{c}}
\newcommand{\Vj}{\mathbf{j}}

\newcommand{\mal}{\cdot}
